\documentclass[sigconf]{acmart}
\newif\ifnotnewacm
\notnewacmfalse

\newif\ifheadnice
\headnicefalse

\usepackage[english]{babel}
\usepackage[utf8]{inputenc}
\usepackage{CJKutf8}

\newcommand{\AUTHORS}{Authors}
\newcommand{\TITLE}{Title}
\newcommand{\KEYWORDS}{Keywords}
\newcommand{\CONFERENCE}{Somewhere}
\newcommand{\COLOR}{yes}
 
\newcommand{\COMMENTS}{yes}

\usepackage{color,balance,xspace,verbatim,ifthen,engord,pifont}

\usepackage{gensymb,amsmath,wasysym,amsthm,marvosym}
\usepackage{cancel}

\usepackage{booktabs,colortbl,diagbox,multirow,tabularx,tablefootnote,circuitikz} 

\usepackage{dblfloatfix} % allow the table* to be place at the page bottom

\usepackage{subcaption} % for subfigure and subtable
\usepackage{graphicx,epsfig,epstopdf,wrapfig}
\graphicspath{{./images/}}
\DeclareGraphicsExtensions{.pdf,.mps,.png,.jpg,.eps,.PNG,.JPG}
\usepackage{threeparttable}
\usepackage{graphicx}
\usepackage{booktabs}

\usepackage{soul} 

\usepackage{algorithm,algorithmic}
\usepackage{url}
\usepackage{tikz}
\newcommand*\circled[1]{\tikz[baseline=(char.base)]{\node[shape=circle,draw,inner sep=0.5pt] (char) {#1};}}
\usepackage{courier} % produce real \texttt

\usepackage{threeparttable}
\usepackage{threeparttablex}
\usepackage{longtable}

\usepackage{enumitem}

\ifthenelse{\equal{\COMMENTS}{yes}}{
	\newcommand{\todo}[1]{\textcolor{red}{\textbf{TODO:} #1}}
	\newcommand{\fye}[1]{\textcolor{red}{#1}}  %content will be excluded
	\newcommand{\remind}[1]{\footnote{\textit{\textcolor{red}{\textbf{Remind:} #1}}}}
	\newcommand{\del}[1]{\color{blue} {\sout{#1}}}
	\newcommand{\p}[1]{\vskip 1ex \noindent\colorbox{yellow}{\parbox{\columnwidth}{#1}}\vskip 4pt}
	\newcommand{\note}[1]{\vskip 4ex \noindent\colorbox{yellow}{\parbox{\columnwidth}{#1}}\vskip 6ex} % highlight
	\newcommand{\q}[1]{\vskip 1ex \noindent\colorbox{magenta}{\parbox{\columnwidth}{\textbf{Question:} #1}}\vskip 4pt} 
	\newcommand{\qa}[1]{\hl{\textbf{Answer:} #1}}
}{
	\newcommand{\todo}[1]{}
	
	\newcommand{\fye}[1]{}
	\newcommand{\remind}[1]{}

	\newcommand{\del}[1]{}
	\newcommand{\p}[1]{}
	\newcommand{\note}[1]{}

	\newcommand{\q}[1]{}
	\newcommand{\qa}[1]{}	
} 

\ifnotnewacm
\ifthenelse{\equal{\COLOR}{yes}}
{\usepackage[colorlinks]{hyperref}} 		% for colored version 
{\usepackage[pdfborder={0 0 0}]{hyperref}} 	% for B&W version
\hypersetup{
	plainpages=false,
	filecolor=cyan,
	linkcolor=red,
	citecolor=black,	
	urlcolor=magenta,
	pdfauthor={\AUTHORS},
	pdftitle={\TITLE},
	pdfsubject={\CONFERENCE},
	pdfkeywords={\KEYWORDS},
	bookmarksopen = {true}	
}
\else
\fi

\ifheadnice
\makeatletter
\renewcommand{\section}{\@startsection{section}{1}{\z@}{-.8ex \@plus -.4ex \@minus -.4ex}{.8ex \@plus .4ex \@minus .4ex}{\normalfont\large\bfseries\MakeTextUppercase}} 
\renewcommand{\subsection}{\@startsection{subsection}{2}{\z@}{-.8ex\@plus -.2ex \@minus -.2ex}{.4ex \@plus .2ex \@minus .2ex}{\normalfont\large\rmfamily\bfseries}} 
\renewcommand{\subsubsection}{\@startsection{subsubsection}{2}{\z@}{-.4ex\@plus -.2ex \@minus -.2ex}{.2ex \@plus .2ex \@minus .2ex}{\normalfont\large\slshape}}

\makeatother
\else
\fi

\graphicspath{{./images/}}

\newif\ifeg
\egfalse

\renewcommand\footnotetextcopyrightpermission[1]{} % removes copyright information 
\setlist[itemize]{leftmargin=10pt,itemsep=0pt,parsep=0pt}
\setlist[enumerate]{leftmargin=0pt}
\usepackage{setspace}
\usepackage{adjustbox}
\usepackage{tcolorbox}
\usepackage{makecell}
\theoremstyle{definition}

\usepackage{pifont}
\usepackage{multirow}
\usepackage{amsmath}
\usepackage{subcaption}

\begin{document}
\fancyhead{} % removes the default page header

\title{AQUILA: A QUIC-Based Link Architecture for \\Resilient Long-Range UAV Communication}

\author{Ximing Huang}
\email{hee@stu.pku.edu.cn}
\affiliation{
  \institution{Peking University}
  \country{China}
  \city{Beijing}
}

\author{Yirui Rao}
\email{rao-yi-rui@email.ncu.edu.cn}
\affiliation{
  \institution{Nanchang University}
  \country{China}
  \city{Nanchang}
}

% \author{Anonymous Authors}
% \email{anonymous@example.edu}
% \affiliation{
%   \institution{Anonymous Institution}
%   \country{Anonymous Country}
%   \city{Anonymous City}
% }

\begin{abstract}
The proliferation of autonomous Unmanned Aerial Vehicles (UAVs) in Beyond Visual Line of Sight (BVLOS) applications is critically dependent on resilient, high-bandwidth, and low-latency communication links. Existing solutions face critical limitations: TCP's head-of-line blocking stalls time-sensitive data, UDP lacks reliability and congestion control, and cellular networks designed for terrestrial users degrade severely for aerial platforms. This paper introduces AQUILA, a cross-layer communication architecture built on QUIC to address these challenges. AQUILA contributes three key innovations: (1) a unified transport layer using QUIC's reliable streams for MAVLink Command and Control (C2) and unreliable datagrams for video, eliminating head-of-line blocking under unified congestion control; (2) a priority scheduling mechanism that structurally ensures C2 latency remains bounded and independent of video traffic intensity; (3) a UAV-adapted congestion control algorithm extending SCReAM with altitude-adaptive delay targeting and telemetry headroom reservation. AQUILA further implements 0-RTT connection resumption to minimize handover blackouts with application-layer replay protection, deployed over an IP-native architecture enabling global operation. Experimental validation demonstrates that AQUILA significantly outperforms TCP- and UDP-based approaches in C2 latency, video quality, and link resilience under realistic conditions, providing a robust foundation for autonomous BVLOS missions.

\end{abstract}

\maketitle

\section{INTRODUCTION}

The operational paradigm of Unmanned Aerial Vehicles (UAVs) is currently shifting from Visual Line-of-Sight (VLOS) flights to Beyond Visual Line-of-Sight (BVLOS) missions. This transition is not merely an incremental upgrade but a critical prerequisite for deploying UAVs in industrial sectors such as large-scale infrastructure monitoring, precision agriculture, and autonomous logistics. However, the feasibility of these advanced operations relies on solving a formidable system-level challenge: establishing a robust communication link capable of reconciling two diametrically opposed traffic requirements. To operate safely, a remote pilot or edge agent requires a high-bandwidth, throughput-oriented video stream for situational awareness, concurrently with an ultra-low-latency, reliability-oriented stream for command and control (C2) telemetry.

Historically, these communication needs were met through direct point-to-point radio links. While effective at short ranges, these proprietary systems are fundamentally constrained by the physics of the radio horizon and signal occlusion, effectively tethering operations to a limited geographic radius. To achieve truly global operability, the industry is increasingly adopting cellular networks (4G/LTE and 5G) as the primary transport substrate~\cite{uav_5g_advances_2019}. Cellular infrastructure promises ubiquitous coverage and unbounded range, theoretically resolving the distance constraints of traditional radio. However, migrating aerial platforms to cellular networks introduces a new class of "hostile" physical layer challenges that standard communication stacks are ill-equipped to handle~\cite{5g_uav_2024}.

\begin{figure}[thpb]
    \centering
    \includegraphics[width=\linewidth]{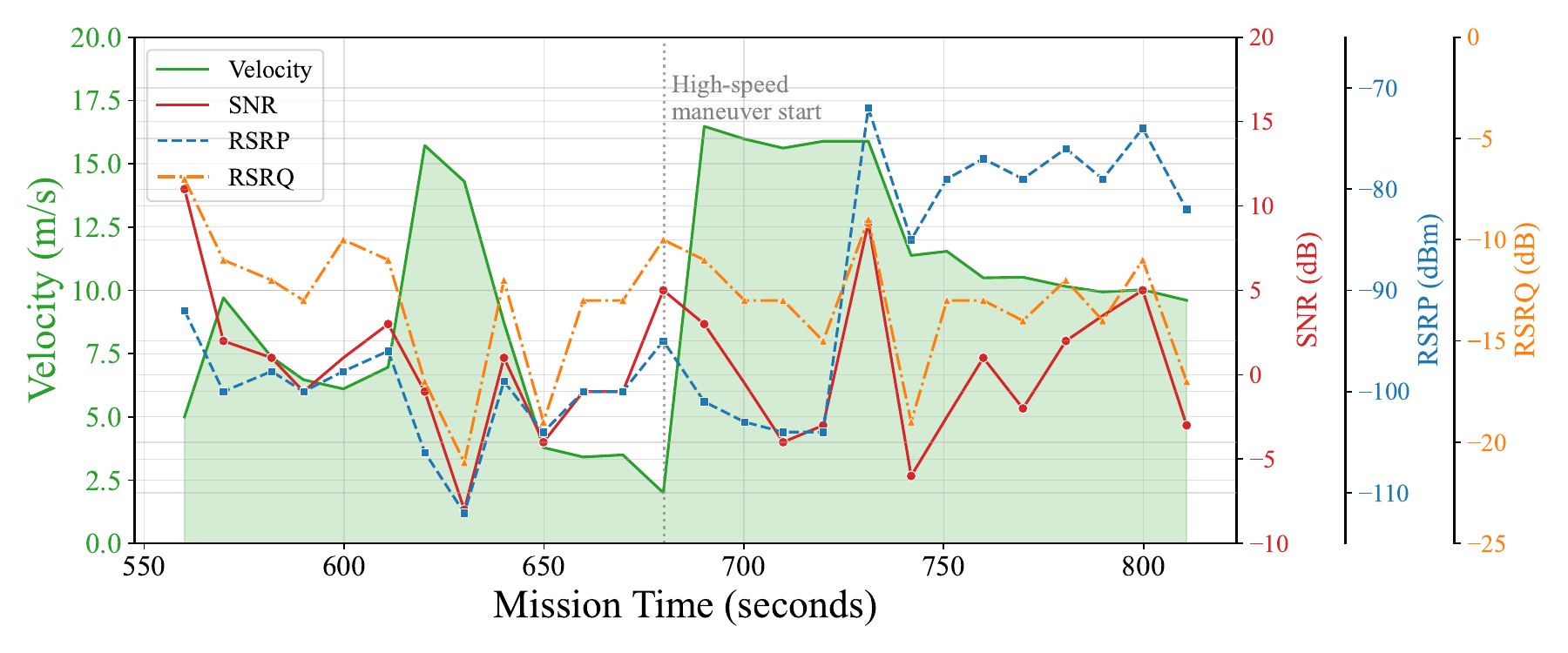}
    \caption{Impact of Aerial Mobility on Link Stability. \textnormal{Field measurements illustrating the correlation between UAV velocity and sharp degradations in SNR, RSRP and RSRQ. Unlike terrestrial links, the aerial channel exhibits extreme volatility due to \textit{side-lobe connectivity} and \textit{inter-cell interference}, fundamental limitations of cellular networks at altitude as identified in 3GPP TR 36.777~\cite{3gpp_tr36777}.}}
    \label{fig:velocity_penalty}
\end{figure}

Cellular networks are architected and optimized for terrestrial users, utilizing antennas that are physically downtilted to serve devices on the ground~\cite{lte_for_uav}. Aerial platforms operating at altitude must connect via the side lobes of these antennas. During flight, the UAV often traverses multiple side lobes rapidly, causing sharp signal changes and leading to a highly unstable connection environment. Field observations confirm that UAVs experience signal quality (RSRQ) fluctuations approaching 20 dB and signal strength (RSRP) variations approaching 60 dBm, alongside significant interference from neighboring cells. Furthermore, the high line-of-sight visibility at altitude exposes the UAV to multiple cell towers simultaneously, triggering frequent and often unnecessary handovers~\cite{handover_challenge}. Consequently, the aerial link is defined not by static bandwidth limitations, but by extreme stochasticity—characterized by rapid jitter, sudden latency spikes, and transient bandwidth collapses.

In this volatile environment, existing communication protocols fail to provide the necessary reliability for safety-critical operations. Commercial off-the-shelf solutions such as \textit{DJI Cellular Dongle Enhanced Transmission} often rely on proprietary implementations that lack interoperability and fail to address the unique physics of the aerial channel~\cite{dji_4g_dongle}. More critically, the standard "hybrid" networking approach—combining Transmission Control Protocol (TCP) for controls and User Datagram Protocol (UDP) for video—suffers from a catastrophic architectural flaw when applied to BVLOS missions~\cite{khuwaja2018survey}.

The root of this failure lies in the independent nature of these protocols. TCP, designed for guaranteed delivery, is susceptible to Head-of-Line (HOL) blocking. In a mixed-traffic environment, if a single TCP segment is lost due to channel noise, the entire stream stalls while waiting for retransmission, delaying time-sensitive C2 commands behind obsolete data~\cite{analysis_for_uav}. Conversely, UDP, commonly used for video, lacks built-in congestion control and reliability. It operates in a "context-blind" manner, unaware of the state of the parallel TCP control link. When the cellular link bandwidth inevitably degrades due to a handover or interference, the UDP video stream continues to push high-bitrate data, flooding the network buffers. This effectively results in a self-inflicted Denial-of-Service (DoS) attack, where the high-volume video traffic starves the critical control link, leading to connection timeouts and a loss of vehicle control~\cite{bufferbloat}. Existing solutions address these challenges in isolation, failing to recognize that the video and control links must be managed as a coupled system.

While industries such as cloud gaming face similar latency-throughput trade-offs and have adopted protocols like WebRTC to mitigate them, these applications fundamentally operate under different assumptions. Cloud gaming relies on stable network topologies supported by ample edge computing resources, where channel variance is minimal. In stark contrast, UAVs operating over LTE utilize a highly volatile physical layer, frequently experiencing precipitous drops in signal quality, alongside resource exhaustion at the base station level. Traditional real-time protocols, optimized for predictable terrestrial channels, exhibit poor resilience under these erratic conditions, rendering them unsuitable for safety-critical BVLOS systems.

To resolve these intertwined challenges, this paper introduces AQUILA, a cross-layer communication architecture explicitly designed for resilient long-range UAV communication. Rather than treating video and telemetry as separate flows competing for resources, AQUILA leverages the QUIC protocol to create a unified transport layer. By utilizing QUIC's multi-stream capabilities~\cite{quic_multiplexed} and the Unreliable Datagram Extension (RFC 9221)~\cite{pauly2022unreliable}, AQUILA multiplexes reliable C2 streams and unreliable video datagrams within a single, encrypted congestion control context. This allows the application to enforce strict priority scheduling at the transport level, ensuring that video traffic serves the mission without ever compromising the integrity of the flight controls.

This paper makes the following specific contributions to the field of resilient cellular communication for high-mobility aerial platforms:

\vspace{-0.5em}

\begin{enumerate}
    \item \textbf{A Unified Transport with Priority Guarantees:} We propose a novel scheduler that maps MAVLink C2 data to reliable QUIC streams and video data to unreliable datagrams. We provide a mathematical stability analysis proving that the latency of the C2 link remains bounded and independent of the video traffic intensity, effectively eliminating head-of-line blocking.
    
    \item \textbf{UAV-Aware Congestion Control:} We introduce a modified version of the Self-Clocked Rate Adaptation for Multimedia (SCReAM) algorithm, integrated directly into the QUIC stack. This implementation features "Altitude-Adaptive Delay Targeting" to distinguish between congestion and handover latency, and a "Telemetry Headroom Reservation" mechanism that throttles video bitrate specifically to preserve bandwidth for control commands.
    
    \item \textbf{Resilient Global Connectivity:} To minimize the "blackout window" during cell tower handovers, AQUILA implements 0-RTT connection resumption with application-layer replay protection. Furthermore, the system is deployed over a transparent, IP-native WireGuard overlay, decoupling the C2 link from physical radio constraints and enabling global peer-to-peer connectivity across NATs.
\end{enumerate}
\section{EXPERIMENTAL PLATFORM}

The AQUILA framework is validated on a custom fixed-wing UAV experimental platform.

\begin{figure}[thpb]
    \centering
    \begin{subfigure}{\linewidth}
        \centering
        \includegraphics[width=\linewidth]{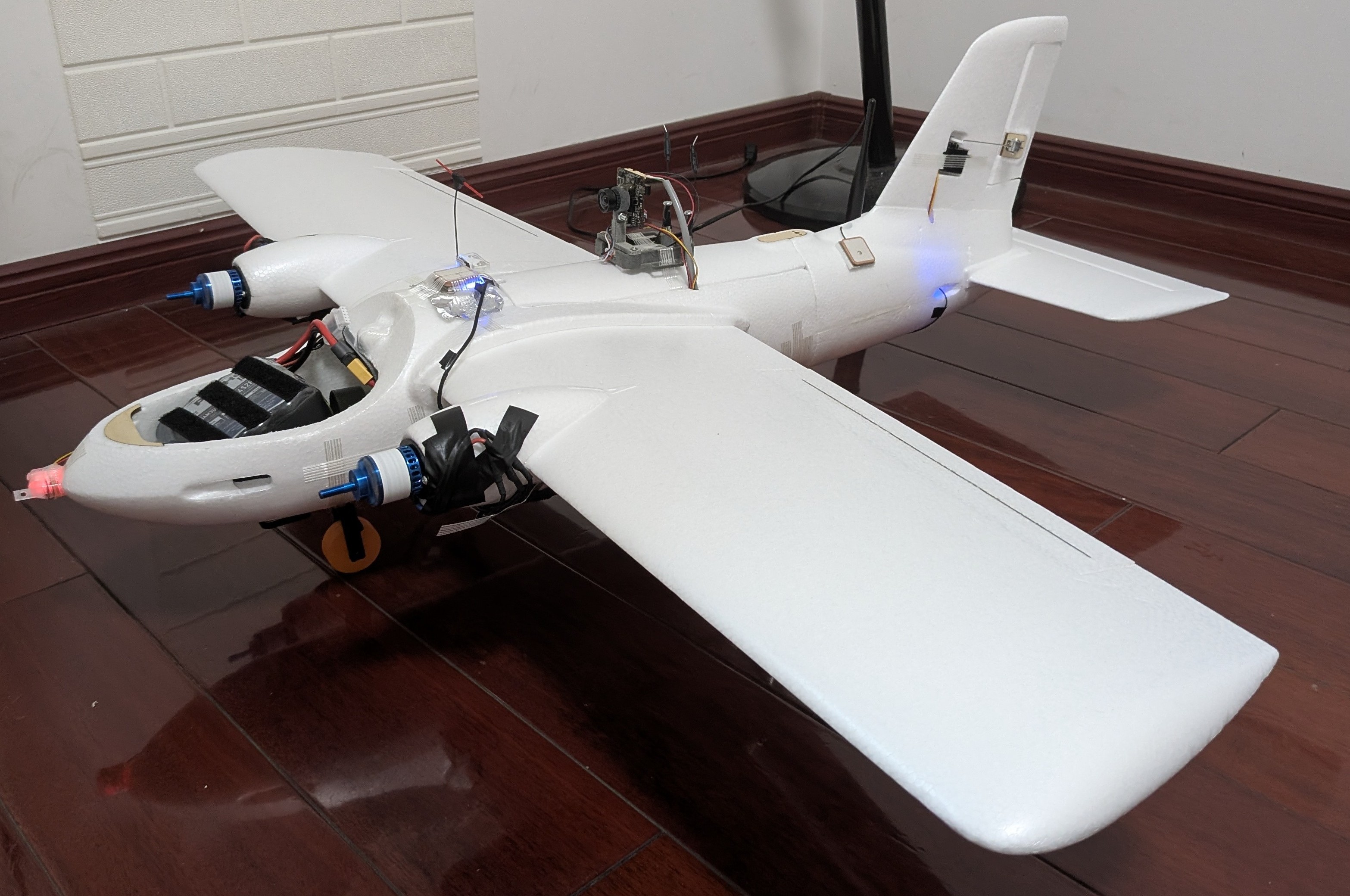}
        \caption{The UAV platform used in this study.}
        \label{fig:aircraft}
    \end{subfigure}
    \begin{subfigure}{\linewidth}
        \centering
        \includegraphics[width=\linewidth]{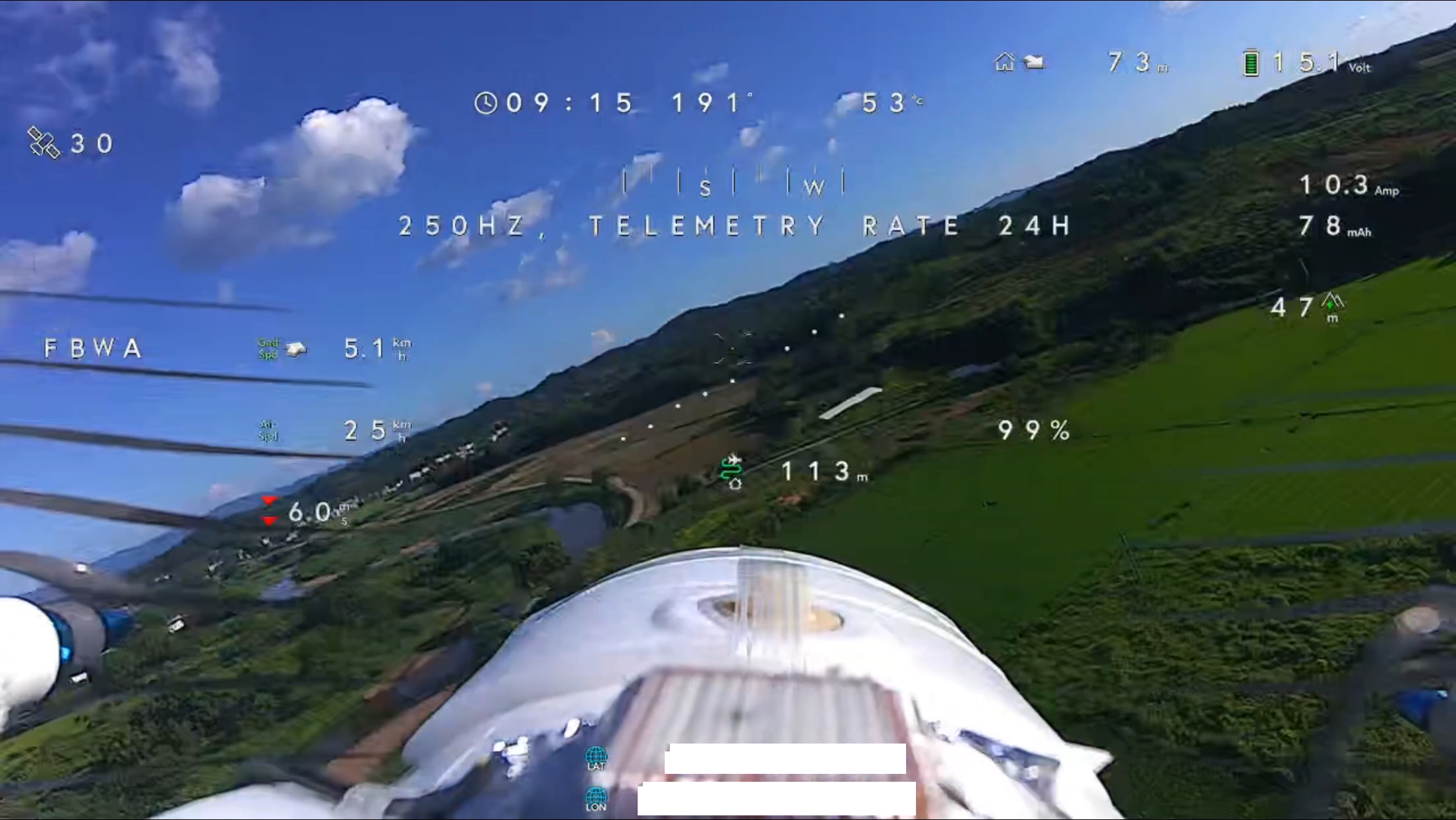}
        \caption{First-Person View (FPV) downlink from the UAV. \textnormal{The on-screen display (OSD) includes key telemetry data such as altitude and speed.}}
        \label{fig:flying}
    \end{subfigure}
    \caption{Experimental testbed}
    \label{fig:testbed}
\end{figure}

\subsection{Hardware Platform}
The platform is built on a durable airframe with a 1100mm wingspan and an 870mm fuselage. Flight control is managed by a Speedybee F405 Wing Mini flight controller running the open-source ArduPilot firmware~\cite{ardupilot}. For optimal performance, the propulsion system is equipped with two Sunnysky X2212 brushless motors, each managed by a Hobbywing Skywalker V2 ESC. The heart of the AQUILA onboard system is an Orange Pi 3B single-board computer, which serves as the companion computer and is essential for concurrently handling the computational load of QUIC encryption and video processing. For video transmission, the platform utilizes a dual-link system composed of a LTE modem and a 5.8 GHz radio. The LTE connectivity is provided by a Quectel EC20CEFAG module configured with diversity antennas: an external omnidirectional monopole antenna as the primary interface, and an embedded Flexible Printed Circuit (FPC) antenna as the secondary diversity receiver.

\begin{figure}[thpb]
  \centering
  \includegraphics[width=\linewidth]{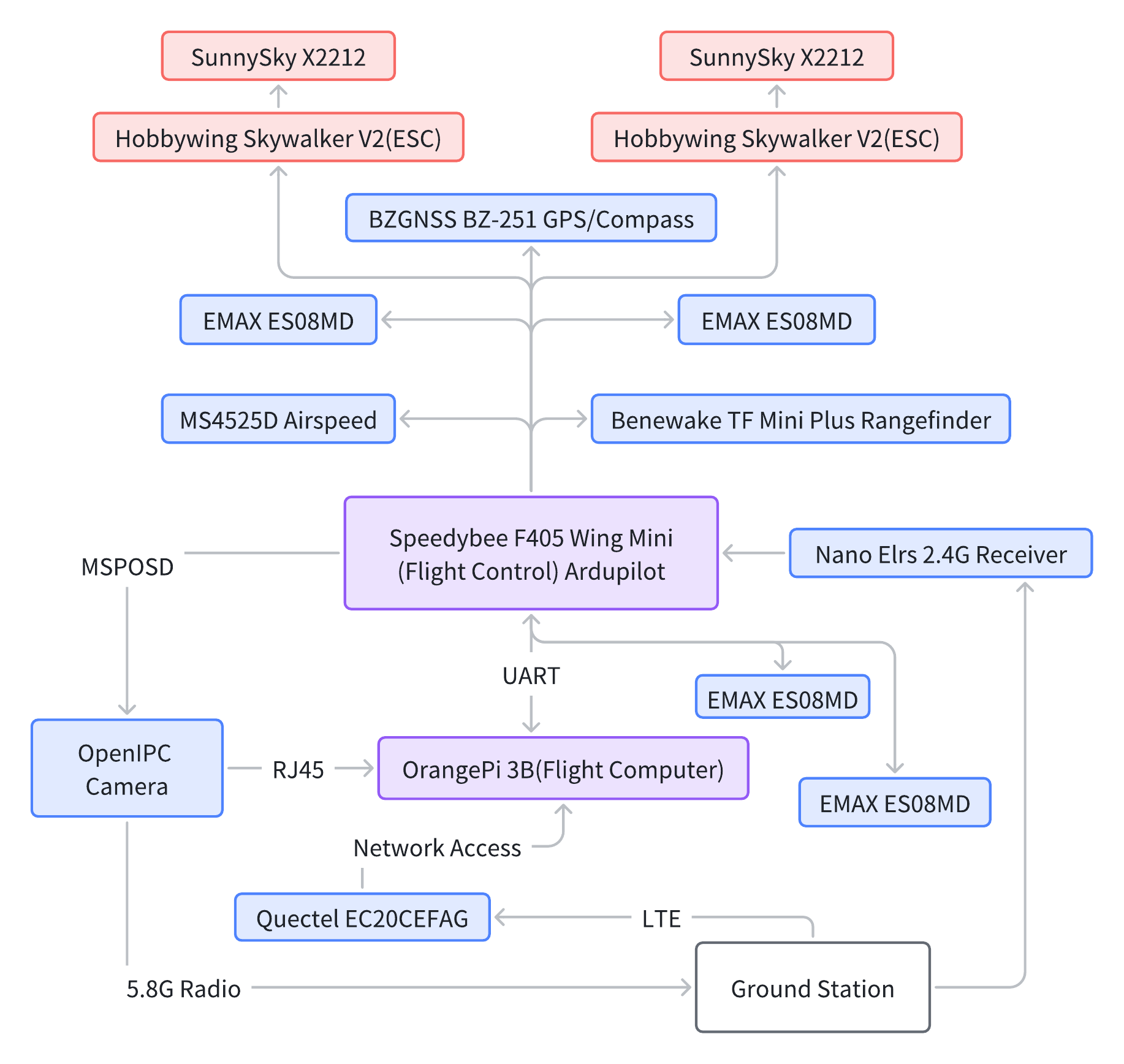}
  \caption{Onboard Hardware Configuration}
  \label{fig:hardware}
\end{figure}

\subsection{Software Implementation and Traffic Scheduling}
\label{subsec:software_impl}

The software architecture is centered on the Orange Pi 3B companion computer, which runs the custom AQUILA application responsible for data aggregation and transport management. The application is developed in Rust to ensure memory safety and high performance, utilizing the \texttt{quiche} library as its core networking component. Its primary function is to process two distinct data flows from onboard sources before transmission:

\begin{itemize}
    \item \textbf{Reliable MAVLink C2 Data Flow:} The Speedybee F405 flight controller, running ArduPilot, continuously generates MAVLink telemetry packets. These packets are serialized and transmitted over a dedicated UART interface to the Orange Pi, where the AQUILA software ingests them for the command and control (C2) channel.
    \item \textbf{Low-Latency Real-Time Video Data Flow:} The OpenIPC IMX307 camera provides the video feed, which is encoded to a H.265 stream and sent to the Orange Pi over UDP. The AQUILA software captures these incoming UDP packets for subsequent transmission.
\end{itemize}

% \paragraph{Concurrency and Scheduling Model.} To effectively manage these flows without introducing latency, the application implements a concurrent queuing model based on Rust's asynchronous runtime. The core function of the scheduler is to explicitly map these logical flows to the appropriate QUIC transport primitives: critical MAVLink packets are routed to \textbf{reliable streams}, while latency-sensitive video frames are encapsulated in \textbf{unreliable datagrams}. We formalize the input buffers as two priority queues: $Q_{high}$ for the MAVLink stream and $Q_{low}$ for the video stream. Let $p_i$ denote a packet arriving at the transport layer at time $t$. The system implements a strict priority scheduling strategy function $f(p_i)$ defined as:

% \begin{equation}
%     f(p_i) = 
%     \begin{cases} 
%     \text{dispatch}(p_i, \text{Stream}_{rel}) & \text{if } p_i \in Q_{high} \\
%     \text{dispatch}(p_i, \text{Dgram}_{unrel}) & \text{if } p_i \in Q_{low} \land \text{IsEmpty}(Q_{high}) \\
%     \text{drop}(p_i) & \text{if } \text{IsFull}(Q_{low})
%     \end{cases}
% \end{equation}
% \label{eq:scheduler}

\paragraph{Concurrency and Scheduling Model.} 
To avoid latency-induced blocking, the application utilizes a concurrent queuing model mapping MAVLink packets to QUIC \textbf{reliable streams} ($Q_{high}$) and video frames to \textbf{unreliable datagrams} ($Q_{low}$). 

Instead of a generic FIFO strategy, we implement the strict priority logic defined in Algorithm~\ref{alg:scheduler}. The scheduler enforces two critical behaviors: (1) \textit{Active Queue Management} at ingestion prevents bufferbloat by dropping excess video frames; (2) \textit{Non-preemptive Dispatch} serves $Q_{high}$ to exhaustion before processing $Q_{low}$. This structure guarantees that telemetry transmission remains independent of video traffic intensity.

\begin{algorithm}[t]
\caption{AQUILA Asynchronous Priority Scheduler}
\label{alg:scheduler}
\begin{algorithmic}[1]
\REQUIRE Incoming Packet Stream $P_{in}$, Network Link Capacity $C(t)$
\ENSURE Prioritized QUIC Transmission
\STATE \textbf{Global} Priority Queues $Q_{high}$ (C2), $Q_{low}$ (Video)

\STATE \textit{// Part 1: Ingestion and Active Queue Management}
\FORALL{packet $p_i \in P_{in}$}
    \IF{$p_i$.type == MAVLink}
        \STATE $Q_{high}$.push($p_i$)
    \ELSIF{$p_i$.type == Video}
        \IF{$Q_{low}$.isFull()}
            \STATE \textbf{Drop}($p_i$) \COMMENT{Prevent bufferbloat}
        \ELSE
            \STATE $Q_{low}$.push($p_i$)
        \ENDIF
    \ENDIF
\ENDFOR

\STATE \textit{// Part 2: Strict Priority Dispatch Loop}
\WHILE{NetworkSocket.isWritable()}
    \IF{\textbf{not} $Q_{high}$.isEmpty()}
        \STATE $p \leftarrow Q_{high}$.pop()
        \STATE \textsc{QuicStreamSend}($p$, Reliable)
    \ELSIF{\textbf{not} $Q_{low}$.isEmpty()}
        \STATE $p \leftarrow Q_{low}$.pop()
        \STATE \textsc{QuicDgramSend}($p$, Unreliable)
    \ELSE
        \STATE \textbf{break} \COMMENT{Yield to async runtime}
    \ENDIF
\ENDWHILE
\end{algorithmic}
\end{algorithm}

\paragraph{Stability Analysis.} We can verify the stability of the C2 link under this scheduler. Let $\lambda_{c2}$ and $\lambda_{vid}$ denote the arrival rates of the C2 and video streams, respectively, and let $\mu$ represent the effective link service rate. In a standard FIFO system (e.g., a single TCP connection), the expected waiting time $E[W]$ is a function of the total aggregate traffic $\rho_{total} = (\lambda_{c2} + \lambda_{vid})/\mu$. Consequently, if $\lambda_{vid}$ surges such that $\rho_{total} \to 1$, the delay for all packets, including C2, approaches infinity ($E[W] \to \infty$).

However, under the AQUILA scheduler defined in Algorithm~\ref{alg:scheduler}, the system behaves as a non-preemptive priority queue. The expected waiting time for the high-priority C2 packets, $E[W_{c2}]$, is decoupled from the video traffic intensity:

\begin{equation}
    E[W_{c2}] \approx \frac{\rho_{c2} \cdot \bar{S}}{1 - \rho_{c2}} + R
    \label{eq:wait_time}
\end{equation}

where $\bar{S}$ is the average service time and $R$ is the residual service time of the packet currently being transmitted. Crucially, Eq.~\ref{eq:wait_time} demonstrates that $E[W_{c2}]$ is independent of $\lambda_{vid}$. Given that MAVLink telemetry consumes negligible bandwidth ($\lambda_{c2} \ll \mu \implies \rho_{c2} \approx 0$), the C2 latency remains minimal and bounded even when the network is heavily congested by video data ($\lambda_{vid} \to \infty$). This provides a theoretical guarantee that the C2 link remains stable regardless of video throughput saturation.

Internally, the ingestion modules act as independent asynchronous tasks producing data into bounded channels. The main event loop, utilizing \texttt{quiche}'s state machine, consumes these channels with the strict priority logic derived above, ensuring that bulk video data never blocks critical C2 telemetry.

\begin{figure*}
    \centering
    \includegraphics[width=\linewidth]{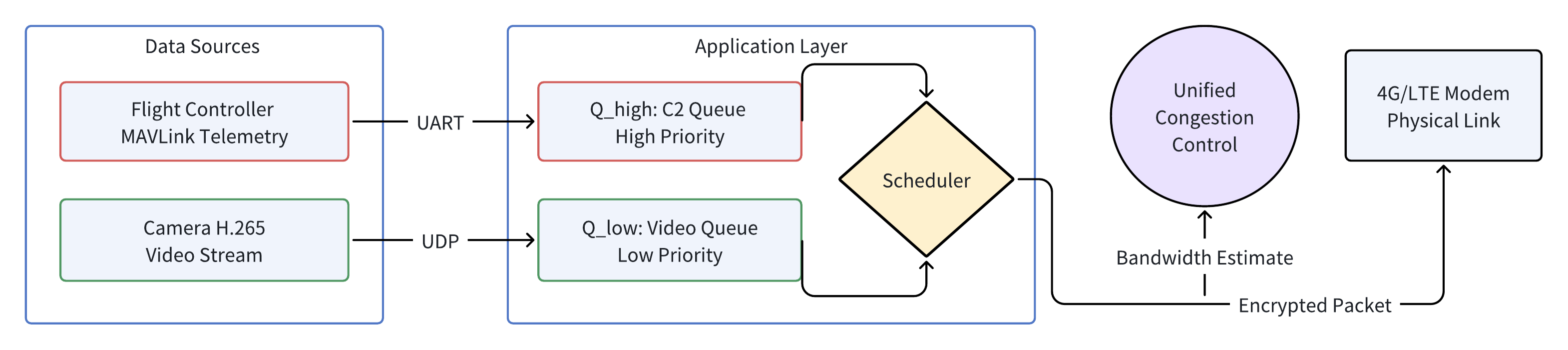}
    \caption{Cross-Layer Data Flow and Scheduling}
    \label{fig:dataflow}
\end{figure*}

\section{PROPOSED COMMUNICATION ARCHITECTURE}

\subsection{Transport Layer Semantics and Dynamics}
\label{subsec:transport_semantics}

While the application-layer scheduler (Algorithm~\ref{alg:scheduler}) enforces input priority, the underlying network behavior is governed by the mapping of these flows to specific QUIC transport primitives. This section analyzes the distinct semantics of these channels and their interaction within a unified cryptographic and congestion context.

\subsubsection*{Reliability Semantics and HOL Elimination:} 
The logical channel denoted as $\text{Stream}_{rel}$ utilizes QUIC's reliable streams. In this mode, the transport layer guarantees ordered delivery through Automatic Repeat reQuest (ARQ), essential for MAVLink integrity. 
Unlike TCP, where a lost packet halts the delivery of all subsequent data (Head-of-Line blocking), QUIC streams are independent. While our implementation uses a single stream for C2, the critical distinction lies in \textit{cross-domain} isolation: a packet loss in the video data (if it were on a stream) or protocol control frames does not block the delivery of C2 telemetry~\cite{pauly2022unreliable}. This ensures that the control loop remains closed even in high-loss environments.

\subsubsection*{Datagram Volatility and Latency:} 
The $\text{Dgram}_{unrel}$ channel leverages the Unreliable Datagram Extension (RFC 9221)~\cite{pauly2022unreliable, togo2021rtp}. This extension allows the transport to treat video frames as ``fire-and-forget'' entities. When the scheduler dispatches a video packet, it is encapsulated in a QUIC DATAGRAM frame. 
Crucially, if this frame is lost in transit, the QUIC stack does not attempt retransmission. This behavior aligns perfectly with the real-time requirement $\tau < \tau_{max}$ established in Section~\ref{subsec:software_impl}, as retransmitting an obsolete video frame typically consumes bandwidth without adding perceptual value.

\subsubsection*{Unified Congestion Dynamics:} 
A fundamental advantage of AQUILA's architecture over traditional hybrid solutions (e.g., separate TCP C2 and UDP Video links) is the \textbf{unified congestion control context}. 
Although C2 and Video possess different reliability semantics, they share a single Congestion Window (CWND). 
When the network bandwidth $C(t)$ drops, the QUIC congestion controller detects the aggregate packet loss or delay from both flows. This allows the transport layer to react atomically—throttling the total send rate—while the application scheduler (Algorithm~\ref{alg:scheduler}) ensures that the reduced bandwidth budget is allocated exclusively to C2 packets first. 
This cross-layer synergy prevents the common ``UDP flood'' scenario where an uncontrolled UDP video stream starves a concurrent, reliable TCP control link, causing connection timeouts.

\subsubsection*{Connection Resilience and Zero-RTT Recovery}

To ensure continuous operability during cellular handovers or NAT rebinding, AQUILA leverages QUIC's connection migration and 0-RTT session resumption features. We formalize the impact of network interruptions on mission safety by defining the \textit{Control Blackout Window} ($W_{cb}$), representing the duration during which the Ground Control Station loses authority over the UAV following a link change:
\begin{equation}
W_{cb} = T_{phy} + T_{handshake}
\end{equation}
where $T_{phy}$ is the physical layer switching delay and $T_{handshake}$ is the protocol renegotiation overhead. For a traditional secure TCP/TLS stack, the handshake overhead requires at least 2 to 3 Round-Trip Times (RTT) before application data can flow ($T_{handshake}^{TCP} \approx 3 \cdot \text{RTT}$). In contrast, AQUILA utilizes cached session tickets to encrypt payload in the very first packet of the new connection, allowing the protocol overhead to approach zero:
\begin{equation}
\lim_{T_{handshake} \to 0} W_{cb}^{AQUILA} \approx T_{phy}
\label{eq:minimize_window}
\end{equation}
This derivation proves that AQUILA minimizes the vulnerability window strictly to the physical layer's limitations.

However, the elimination of the handshake introduces a security trade-off: 0-RTT data is theoretically susceptible to replay attacks. AQUILA addresses this challenge through a cross-layer defense strategy. While the transport layer focuses on minimizing latency, the application layer ensures integrity. The MAVLink protocol embeds monotonic sequence numbers and timestamps in each command packet. Consequently, even if a malicious actor replays a captured 0-RTT C2 packet, the flight controller's parser will identify the payload as duplicate or obsolete and silently discard it. This ensures that the system benefits from instant link recovery without compromising command authority.

\subsection{Context-Aware Congestion Control}
\label{subsec:scream_cca}

While Section~\ref{subsec:transport_semantics} establishes the transport semantics, effective operation in dynamic cellular environments requires a congestion control algorithm (CCA) specifically tailored for low-latency media. Standard loss-based algorithms (e.g., CUBIC~\cite{cubic}) operate strictly at the transport layer, often leading to a mismatch between the video encoder's output rate and the network's capacity~\cite{self-clocked_rate}.

To address this, we integrated and adapted the \textit{Self-Clocked Rate Adaptation for Multimedia} (SCReAM) algorithm~\cite{RFC8298} into our Rust-based QUIC stack. However, the standard SCReAM implementation assumes relatively stable terrestrial links~\cite{scream_in_5g, bit_adapt}. For high-mobility aerial platforms, we introduced two UAV-specific enhancements: \textbf{Altitude-Adaptive Delay Targeting} and \textbf{Telemetry Headroom Reservation}. Throughout the rest of this paper, we denote this optimized, aerial-focused variant as \texttt{SCReAM-FPV}.

\subsubsection*{Adaptive Queue Delay Estimation}
AQUILA prioritizes latency stability over maximum throughput. Standard SCReAM uses a fixed queue delay target. However, our field tests indicated that UAVs at altitude experience drastic shifts in baseline latency due to cell tower handovers and side-lobe interference. A static target often misinterprets these structural latency shifts as congestion.

To mitigate this, we implemented an adaptive queuing delay estimator. Let $RTT(t)$ denote the smoothed round-trip time. We define the estimated queuing delay $\hat{d}_q(t)$ relative to a sliding window minimum, but we dynamically scale the tolerance threshold:

\begin{equation}
    \hat{d}_q(t) = RTT(t) - \min_{k \in [t-W, t]} (RTT(k))
    \label{eq:queue_delay}
\end{equation}

where $W$ is the observation window (20s). Crucially, unlike the standard implementation, the target delay threshold $d_{target}$ is not constant but follows the baseline link characteristics:
\begin{equation}
    d_{target}(t) = \max(d_{base}, \ \kappa \cdot \min_{k \in [t-W, t]} (RTT(k)))
\end{equation}
Here, $\kappa$ is a tolerance factor (set to $1.5$ in experimentation) that prevents the window from collapsing during the inevitable latency spikes caused by LTE handovers.

\subsubsection*{Congestion Window Dynamics}
Based on $\hat{d}_q(t)$, the transport layer adjusts the congestion window ($cwnd$). Our implementation follows a delay-driven update law designed to keep the bottleneck queue occupancy below the adaptive $d_{target}(t)$. The state transition is formalized as:

\begin{equation}
    cwnd(t+1) = 
    \begin{cases} 
    cwnd(t) + \frac{S}{cwnd(t)} & \text{if } \hat{d}_q(t) \le d_{target}(t) \\
    cwnd(t) \cdot (1 - \beta \frac{\hat{d}_q(t) - d_{target}(t)}{\hat{d}_q(t)}) & \text{if } \hat{d}_q(t) > d_{target}(t)
    \end{cases}
    \label{eq:cwnd_dynamics}
\end{equation}

where $S$ is the packet size and $\beta$ is a multiplicative decrease factor. This ensures the system probes for bandwidth aggressively when safe but reacts proportionally to delay violations.

\subsubsection*{Safety-Critical Rate Coupling}
A critical innovation in AQUILA is the strict coupling between the transport layer and the video encoder, modified to guarantee C2 link survival. 

In standard SCReAM, the target bitrate is derived directly from the estimated throughput. In our UAV architecture, we enforce a \textbf{Telemetry Headroom Reservation}. The effective transmission rate $R_{tx}(t)$ is first reduced by a safety margin $R_{safe}$ (reserved for MAVLink traffic) before being fed back to the video encoder:

\begin{equation}
    R_{enc}(t) = \max(R_{min}, \min(R_{max}, \gamma \cdot (R_{tx}(t) - R_{safe})))
    \label{eq:rate_coupling}
\end{equation}

Here, $R_{safe}$ is statically configured based on the MAVLink stream bandwidth (approx. 15 kbps), and $\gamma$ is a damping factor. This equation ensures that the generated video stream never consumes the entirety of the channel capacity, fundamentally eliminating head-of-line blocking for control commands at the application buffer level.

Furthermore, we leverage SCReAM's Credit Mechanism. When the video stream is throttled by Eq.~\ref{eq:rate_coupling}, the transport accumulates transmission credits. These credits allow high-priority MAVLink packets to bypass the congestion window restrictions immediately, ensuring that control commands ($Stream_{rel}$) maintain near-zero latency even when the video link is saturated.

\subsection{Global Connection via Secure Overlay Network}
To achieve global operational range beyond the limitations of local radio links, AQUILA implements a fully IP-native control plane by leveraging WireGuard to create a secure overlay network~\cite{donenfeld2017wireguard}. This approach abstracts the underlying cellular carrier NATs, assigning stable, private IP addresses to both the UAV and the Ground Control Station (GCS). This enables a direct peer-to-peer topology where the C2 link operates as if on a local network, ensuring addressability regardless of physical location.

\paragraph{Cryptographic Optimization} While the overlay network ensures reachability, encapsulating QUIC within a VPN tunnel introduces potential redundancy. Since WireGuard already provides robust mutual authentication and encryption, applying full TLS validation at the transport layer incurs unnecessary computational overhead. To mitigate this ''double encryption'' cost, we optimized the AQUILA stack by explicitly disabling the certificate verification process within the QUIC handshake, effectively offloading anti-replay defenses to the WireGuard tunnel. The system relies on the strict trust boundary established by the WireGuard interface, allowing the transport layer to focus solely on congestion control without duplicating cryptographic verification steps.

\paragraph{MTU Adaptation} A critical challenge identified during experimental deployment was the mismatch in Maximum Transmission Unit (MTU) sizes. The default MTU configuration of the WireGuard interface was found to be smaller than the standard QUIC packet size. Without intervention, this discrepancy forces the network stack to fragment QUIC packets at the overlay layer, which is detrimental to real-time performance. To address this, we implemented a strict MTU clamping strategy. We adjusted the WireGuard interface parameters and simultaneously configured the QUIC stack's \texttt{max\_datagram\_size} to ensure that its payload fits precisely within the overlay's effective payload capacity, guaranteeing that video datagrams are transmitted atomically.

Safety is paramount: if the link fails entirely, the ArduPilot firmware's GCS Failsafe is triggered, commanding an autonomous Return-to-Launch (RTL).

\section{EXPERIMENTAL METHODOLOGY}

A rigorous experimental evaluation was conducted to validate the performance of the AQUILA architecture.

\subsection{Field Test Data Collection}
To establish a realistic baseline for our network emulation, we conducted a series of field experiments in city and suburban environments characterized by moderate cellular tower density, with results shown in Appendix Figure~\ref{fig:fieldtest}. Preliminary experiments involving complex horizontal trajectories revealed that variations in horizontal distance exerted negligible influence on signal quality compared to altitude changes. Consequently, to strictly isolate these altitude-dependent effects (e.g., side-lobe interference~\cite{3gpp_tr36777}) from horizontal path loss and shadowing, we adopted a vertical profiling strategy. Given the forward-flight constraints of our fixed-wing platform, the UAV performed helical ascents from ground level to a maximum altitude of 500~m, maintaining a tight loitering radius to constrain horizontal displacement.

We acknowledge that the banking angle (typically 15--20$^{\circ}$) required for such maneuvers introduces periodic variations in antenna orientation relative to the cell tower. However, the UAV is equipped with a Quectel EC20CEFAG module utilizing a dual-antenna diversity scheme: an external vertically polarized omnidirectional monopole antenna as the primary interface, and an embedded Flexible Printed Circuit (FPC) antenna for diversity reception. This configuration provides a toroidal radiation pattern with a broad vertical beamwidth, ensuring a robust link budget even during banking maneuvers. Thus, the signal fluctuations captured in our traces represent genuine aerial channel stochasticity rather than experimental artifacts.

The collected measurement data, visualized in Fig.~\ref{fig:fieldtest}, reveal the stochastic nature of the aerial channel. Notably, across all tested altitudes, the packet loss rate was consistently measured at 0\%. This phenomenon is attributed to the underlying LTE physical layer's robust Modulation and Coding Scheme (MCS) and Hybrid ARQ mechanisms, which prioritize delivery reliability at the expense of latency. Consequently, signal degradation manifested not as packet drops, but as significant jitter and latency spikes.

However, simply replaying a specific flight trace cannot capture the full spectrum of edge cases required for a robust evaluation, nor does it allow for the isolation of specific variables. Therefore, rather than using the raw traces directly as inputs, we utilized these empirical insights to formulate a set of representative emulation scenarios. By configuring deterministic constraints derived from the lower-bounds of our field measurements (e.g., a sustained 3~Mbps bandwidth cap combined with 120~ms round-trip delay), we established a controlled environment. This approach allows us to rigorously stress-test AQUILA against the specific failure modes---such as bandwidth starvation and bufferbloat---identified during the physical flights, while ensuring strict repeatability as detailed in Section~\ref{subsec:net_emulation}.

\subsection{Controlled Environment: Network Emulation}
\label{subsec:net_emulation}
Field validation, while essential, lacks the controllability required for comparative analysis due to stochastic variables like base station handovers and background traffic. To ensure strict repeatability and a fair comparison, we established a deterministic emulation environment on Microsoft Azure Standard D4ds v4 instances (4 vCPUs, 16 GiB RAM). We utilized the Mahimahi suite~\cite{mahimahi} for trace-driven emulation, converting field throughput measurements into execution traces. By layering \texttt{mm-link} to enforce bandwidth constraints and \texttt{mm-delay} to inject latency spikes, the testbed faithfully reproduces the hostile aerial channel dynamics in a controlled setting.

\paragraph{Baseline Protocols Implementation and Fairness}
Baselines were implemented using the industry-standard \texttt{GStreamer} framework for SRT, WebRTC, and RTP to minimize implementation bias. For KCP, we integrated a high-performance Rust implementation (via \texttt{crates.io/crates/kcp}). We posit that the observed performance gaps—specifically KCP's congestion collapse and WebRTC's service failure—stem from structural limitations in their congestion control designs rather than parametric tuning. Therefore, we maintain that further micro-optimization of the baselines would not alter the qualitative conclusions of this study.

\subsection{Performance Metrics and Assessment}

To rigorously evaluate AQUILA's performance against the BVLOS requirements defined in Section 1, we established a multi-dimensional assessment framework focusing on perceptual quality, signal fidelity, transport latency, and link resilience.

\paragraph{Video Quality Assessment}
To capture a comprehensive view of video transmission performance, we utilized a triad of quality metrics ranging from signal fidelity to perceptual experience:

\begin{itemize}
    \item Perceptual Quality (VMAF): We employed Netflix's \textit{Video Multimethod Assessment Fusion} (VMAF) as our primary metric~\cite{rassool2017vmaf}. VMAF fuses spatial and temporal features to model human visual perception, providing a high correlation with subjective Mean Opinion Scores (MOS).
    
    \item Signal Fidelity (PSNR \& SSIM): To provide a baseline comparison against traditional literature, we calculated the \textit{Peak Signal-to-Noise Ratio} (PSNR) and the \textit{Structural Similarity Index Measure} (SSIM). PSNR quantifies the reconstruction quality at the pixel level, while SSIM evaluates the degradation of structural information (luminance, contrast, and structure)~\cite{psnrssim}.
\end{itemize}

All metrics were computed by comparing the received frames $V_{rx}$ against the source reference $V_{src}$ using FFmpeg's statistical filters.

\paragraph{Frame Delivery Latency}
We define latency as the time interval between the completion of frame sending and its availability for display. To ensure alignment in the presence of packet loss and reordering, we embed a timestamp (frame sequence ID $seq_i$) directly into the source video frames. Upon decoding during post-processing, the sequence ID is extracted to calculate the latency $L_{vid}$ for the $i$-th frame:
\begin{equation}
    L_{vid}(i) = t_{rx}(i) - t_{tx}(i)
\end{equation}
where $t_{tx}$ is the embedded sending timestamp and $t_{rx}$ is the reception timestamp. This metric reflects the pure performance of each protocol.

\paragraph{C2 Link Integrity}
For the safety-critical command and control link, reliability is paramount. We calculated the Packet Loss Ratio ($PLR_{c2}$) by tracking gaps in the monotonic MAVLink sequence numbers embedded in each packet header. For a transmission window of $N$ expected packets with $M$ unique received packets, the loss is defined as:
\begin{equation}
    PLR_{c2} = 1 - \frac{M}{N}
\end{equation}
This allows for the precise detection of even single-packet losses essential for evaluating control authority stability.
\section{RESULTS AND ANALYSIS}

\begin{table*}[t]
\centering
\small
\renewcommand{\arraystretch}{1.1}
\setlength{\tabcolsep}{0pt}

\begin{tabular*}{\textwidth}{@{\extracolsep{\fill}} l l c c c c c c c c }
\toprule
\multirow{2}{*}{\textbf{Scenario}} & \multirow{2}{*}{\textbf{Metric}} & \multicolumn{3}{c}{\textbf{QUIC Variants}} & \multicolumn{5}{c}{\textbf{Traditional Real-Time Protocols}} \\
\cmidrule(lr){3-5} \cmidrule(l){6-10}
 & & \textbf{SCReAM-FPV} & SCReAM & BBR & SRT & RTP & RTP w/ 10\% FEC & KCP & WebRTC \\
\midrule

% ==========================================
% Scenario 1
% ==========================================
\multirow{4}{*}{\parbox{2.8cm}{\raggedright \textbf{1. Real-world Trace} \\ \scriptsize (Verizon LTE Driving Data from Mahimahi repo)}} 
 & Latency (ms) & 212.9 & 483.9 & 394.7 & N/A & N/A & \textbf{23.3} & 793.1 & N/A \\
 & VMAF         & \textbf{92.20} & 56.68 & 83.83 & 0.22 & 0.02 & 0.51 & 43.38 & 0.22 \\
 & SSIM         & \textbf{0.976} & 0.885 & 0.951 & 0.644 & 0.632 & 0.507 & 0.899 & 0.648 \\
 & PSNR (dB)    & 24.07 & \textbf{25.28} & 21.02 & 11.59 & 11.57 & 8.98 & 20.41 & 11.70 \\
\midrule

% ==========================================
% Scenario 2
% ==========================================
\multirow{4}{*}{\parbox{2.8cm}{\raggedright \textbf{2. BW Constrained} \\ \scriptsize (3Mbps, 120ms)}} 
 & Latency (ms) & 580.3 & 593.4 & 445.5 & N/A & N/A & \textbf{199.0} & 3671.0 & N/A \\
 & VMAF         & \textbf{81.16} & 75.31 & 6.50 & 0.01 & 0.54 & 0.02 & 39.75 & 0.00 \\
 & SSIM         & \textbf{0.954} & 0.933 & 0.673 & 0.639 & 0.652 & 0.622 & 0.883 & 0.632 \\
 & PSNR (dB)    & \textbf{22.55} & 19.95 & 12.59 & 11.65 & 11.70 & 11.25 & 20.56 & 11.58 \\
\midrule

% ==========================================
% Scenario 3
% ==========================================
\multirow{4}{*}{\parbox{2.8cm}{\raggedright \textbf{3. High Latency} \\ \scriptsize (Unlimited Bandwidth, 120ms)}} 
 & Latency (ms) & \textbf{128.4} & 128.6 & 320.3 & 137.7 & 137.0 & 137.4 & 140.8 & 137.6 \\
 & VMAF         & 99.06 & 99.08 & 80.52 & \textbf{100.0} & \textbf{100.0} & \textbf{100.0} & 99.91 & \textbf{100.0} \\
 & SSIM         & 1.000 & 1.000 & 0.961 & \textbf{1.000} & \textbf{1.000} & \textbf{1.000} & 1.000 & \textbf{1.000} \\
 & PSNR (dB)    & 57.19 & 59.39 & 26.31 & $\infty$ & $\infty$ & $\infty$ & 49.70 & $\infty$ \\
\midrule

% ==========================================
% Scenario 4
% ==========================================
\multirow{4}{*}{\parbox{2.8cm}{\raggedright \textbf{4. Severe Constraint} \\ \scriptsize (1Mbps, 20ms)}} 
 & Latency (ms) & 1776.8 & 1769.3 & \textbf{1753.8} & N/A & N/A & N/A & 10173.8 & N/A \\
 & VMAF         & \textbf{39.85} & 0.81 & 1.41 & 0.00 & 0.00 & 0.00 & 39.51 & 0.00 \\
 & SSIM         & 0.769 & 0.443 & 0.653 & 0.633 & 0.633 & 0.624 & \textbf{0.887} & 0.633 \\
 & PSNR (dB)    & 13.49 & 7.39 & 12.15 & 11.63 & 11.63 & 11.39 & \textbf{20.54} & 11.64 \\
\bottomrule
\end{tabular*}

\vspace{0.5ex}
\parbox{\textwidth}{\footnotesize 
\textbf{Note:} \textbf{Bold} indicates the best performance in each metric. \textbf{N/A} denotes effective service failure where valid latency metrics could not be derived due to severe frame loss or decoding artifacts preventing timestamp extraction. \\}
\caption{Video Transmission Performance Comparison across Protocols and Scenarios}
\label{tab:full_comparison}
\end{table*}

To comprehensively evaluate the proposed video transmission architecture, we analyzed its performance against a wide range of industry-standard protocols. The benchmarks included standard QUIC congestion controls (BBR, original SCReAM), as well as established real-time transport protocols including SRT, WebRTC, RTP (with and without FEC), and KCP. The comparative results across four distinct network scenarios are summarized in Table~\ref{tab:full_comparison}.

\subsection{Analysis of Video Transmission Performance}
\label{subsec:video_quality}

The performance analysis reveals that our optimized QUIC using \texttt{SCReAM-FPV} consistently outperforms baseline approaches, particularly in dynamic and highly constrained network environments.

\paragraph{Performance in Dynamic Real-world Networks}
Scenario 1 utilizes a real-world Verizon LTE trace, characterized by rapid bandwidth fluctuations and jitter. Here, the advantage of the proposed \texttt{SCReAM-FPV} is most prominent. It achieves a high VMAF score of 92.20 and an SSIM of 0.976, significantly surpassing the original SCReAM (VMAF 56.68) and BBR (VMAF 83.83). Notably, traditional protocols such as SRT, WebRTC, and RTP failed to sustain a serviceable video stream, resulting in near-zero VMAF scores and unrecognizable frames (marked as N/A for latency). While \texttt{RTP (w/ 10\% FEC)} achieved the lowest latency (23.3 ms), it did so at the cost of catastrophic quality loss (VMAF 0.51), rendering the video unusable. In contrast, our approach strikes an optimal balance, maintaining high visual fidelity with a manageable latency of 212.9 ms.

\paragraph{Resilience Under Bandwidth Constraints}
Scenarios 2 and 4 evaluate the protocols under static bandwidth limitations. In Scenario 2 (3 Mbps, 120 ms delay), \texttt{SCReAM-FPV} again leads with a VMAF of 81.16. Standard BBR struggles significantly here, dropping to a VMAF of 6.50, demonstrating its inability to adapt to the combination of high delay and limited bandwidth. The robustness of our architecture is best demonstrated in Scenario 4, the "Severe Constraint" case (1 Mbps, 20 ms delay). This scenario represents a critical bottleneck where most protocols collapse. As shown in Table~\ref{tab:full_comparison}, \texttt{QUIC (SCReAM-FPV)} is the \textit{only} protocol capable of delivering a recognizable video feed, maintaining a VMAF of \textbf{39.85}. All other baselines exhibit total service failure (VMAF $\approx$ 0--1.4). Although the latency increases to 1.7s due to the extreme bottleneck, the proposed method preserves situational awareness where others result in a blackout.

\paragraph{Efficiency in Unconstrained Environments}
Scenario 3 (High Latency, Unconstrained Bandwidth) serves as a baseline check to ensure the proposed optimizations do not introduce unnecessary overhead when network resources are abundant. The results indicate that \texttt{QUIC (SCReAM-FPV)} achieves near-perfect quality (VMAF 99.06), performing on par with SRT and WebRTC (VMAF 100.00). Furthermore, the measured latency of 128.4 ms is remarkably close to the physical one-way delay of 120 ms, confirming that the protocol introduces negligible processing delay under ideal conditions.

\paragraph{Instability of Aggressive ARQ (KCP)}
Distinct performance characteristics were observed with KCP, which employs an aggressive Automatic Repeat reQuest (ARQ) mechanism optimized for low latency. While KCP matches state-of-the-art protocols in unconstrained environments (Scenario 3, Latency: 140.8 ms), it exhibits extreme instability under bandwidth constraints. As detailed in Table~\ref{tab:full_comparison}, in Scenario 4 (1 Mbps), KCP maintains a relatively high structural similarity (SSIM 0.887) but at the cost of unacceptable latency (10.17 s). This phenomenon attributes to a \textit{congestion collapse} caused by the protocol's aggressive retransmission logic. When the network bandwidth is saturated, KCP's fast-retransmit mechanism interprets queuing delay as packet loss, triggering redundant retransmissions. These retransmissions exacerbate the bottleneck, creating a positive feedback loop that effectively destroys real-time interactivity. This indicates that without an adaptive back-off, aggressive ARQ protocols are prone to entering a metastable state where high reliability is traded for infinite latency spikes.

\paragraph{Fragility of Delay-Based Control (WebRTC)}
Conversely, WebRTC demonstrated unexpected fragility in these bandwidth-constrained scenarios, yielding negligible VMAF scores despite being an industry benchmark. This behavior stems from the design philosophy of its default Google Congestion Control (GCC). GCC is a delay-based controller that aggressively throttles sending rates upon detecting queuing delay gradients. In the tested low-bandwidth environments, the available throughput fell below the minimum bitrate required to encode the high-motion video content effectively. Consequently, the WebRTC stack prioritized latency minimization by aggressively skipping frames and reducing resolution to non-viable levels, resulting in effective service failure.

\paragraph{Summary of Comparative Analysis}
In summary, the comparative analysis highlights the structural limitations of existing protocols in challenging aerial networks. Traditional protocols exhibit a dichotomy of failure: aggressive protocols like KCP cause congestion collapse (high latency), while conservative protocols like WebRTC cause service starvation (low quality). The \texttt{QUIC (SCReAM-FPV)} approach demonstrates superior adaptability by effectively navigating this trade-off, ensuring video continuity and acceptable quality even in the presence of severe real-world network impairments.

\subsection{Command and Control (C2) Link Integrity}
\label{subsec:c2_integrity}

For the C2 link, reliability is paramount. We compared AQUILA's use of QUIC reliable streams against a standard TCP implementation. The key metrics were C2 packet loss and round-trip latency.

\begin{figure}[thpb]
    \centering
    \includegraphics[width=\linewidth]{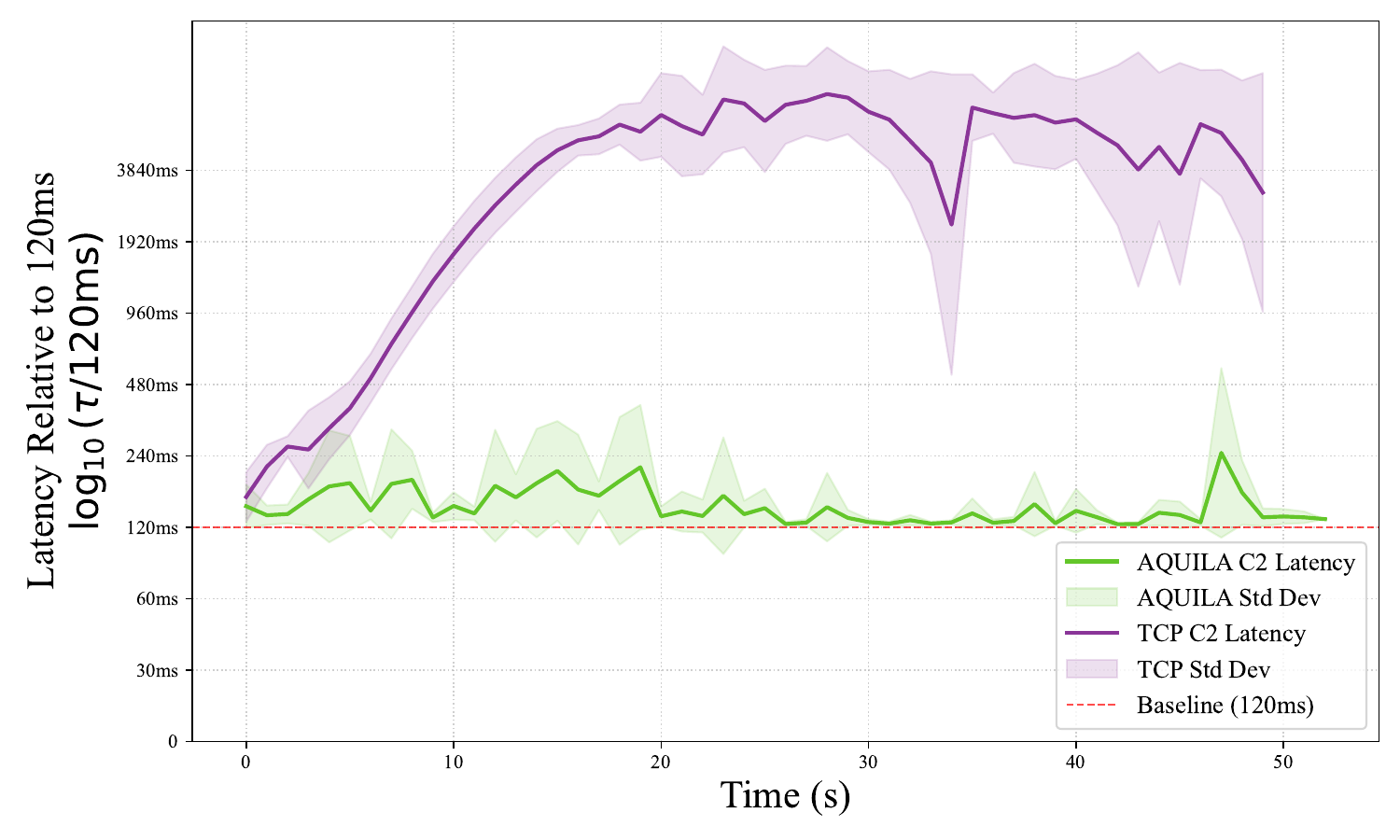}
    \caption{C2 Latency Stability Comparison under Dynamic Network Conditions. \textnormal{The plot illustrates the logarithmic latency ($\log_{10}$) of C2 packets relative to a 100ms baseline over time. TCP (purple) exhibits exponential latency growth due to retransmissions and Head-of-Line blocking when the network is congested. In contrast, AQUILA (green) maintains a bounded latency profile near the baseline, empirically validating the priority scheduling mechanism's ability to decouple C2 telemetry from background video traffic intensity. Error bars represent the standard deviation across 5 experimental trials.}}
    \label{fig:c2}
\end{figure}

As shown in Figure~\ref{fig:c2}, the raw UDP baseline is unsuitable for C2 transmission, as its packet loss is directly proportional to the network congestion. Both AQUILA and TCP guarantee near-zero packet loss due to their reliability mechanisms. However, under the congested scenario, TCP's latency increases significantly due to retransmissions and potential head-of-line blocking. AQUILA demonstrates a clear advantage by providing the same level of reliability as TCP but with a substantially lower latency penalty, ensuring that critical control commands are delivered both reliably and promptly.

\subsection{Traffic Coexistence and Scheduler Dynamics}
\label{sec:traffic_coexistence}

While the previous sections analyzed the video and control links in isolation, the primary contribution of AQUILA is its ability to manage these conflicting flows within a unified transport context. To validate the \textit{Safety-Critical Rate Coupling} mechanism proposed in Eq.~\ref{eq:rate_coupling}, we conducted a bandwidth stress test designed to force the two traffic types to compete for strictly limited resources.

\paragraph{Headroom Reservation Verification}
In this experiment, the UAV transmits a high-bitrate video stream alongside a continuous 10 Hz MAVLink C2 stream. We utilized the network emulator to impose a drastic bandwidth restriction, simulating a "bandwidth collapse" scenario typical of aerial cell-edge conditions. The physical link capacity $C(t)$ is initially set to 5 Mbps (simulating strong LTE coverage) and is abruptly throttled to 1 Mbps at $t=10s$.

\begin{figure}[thpb]
    \centering
    \includegraphics[width=\linewidth]{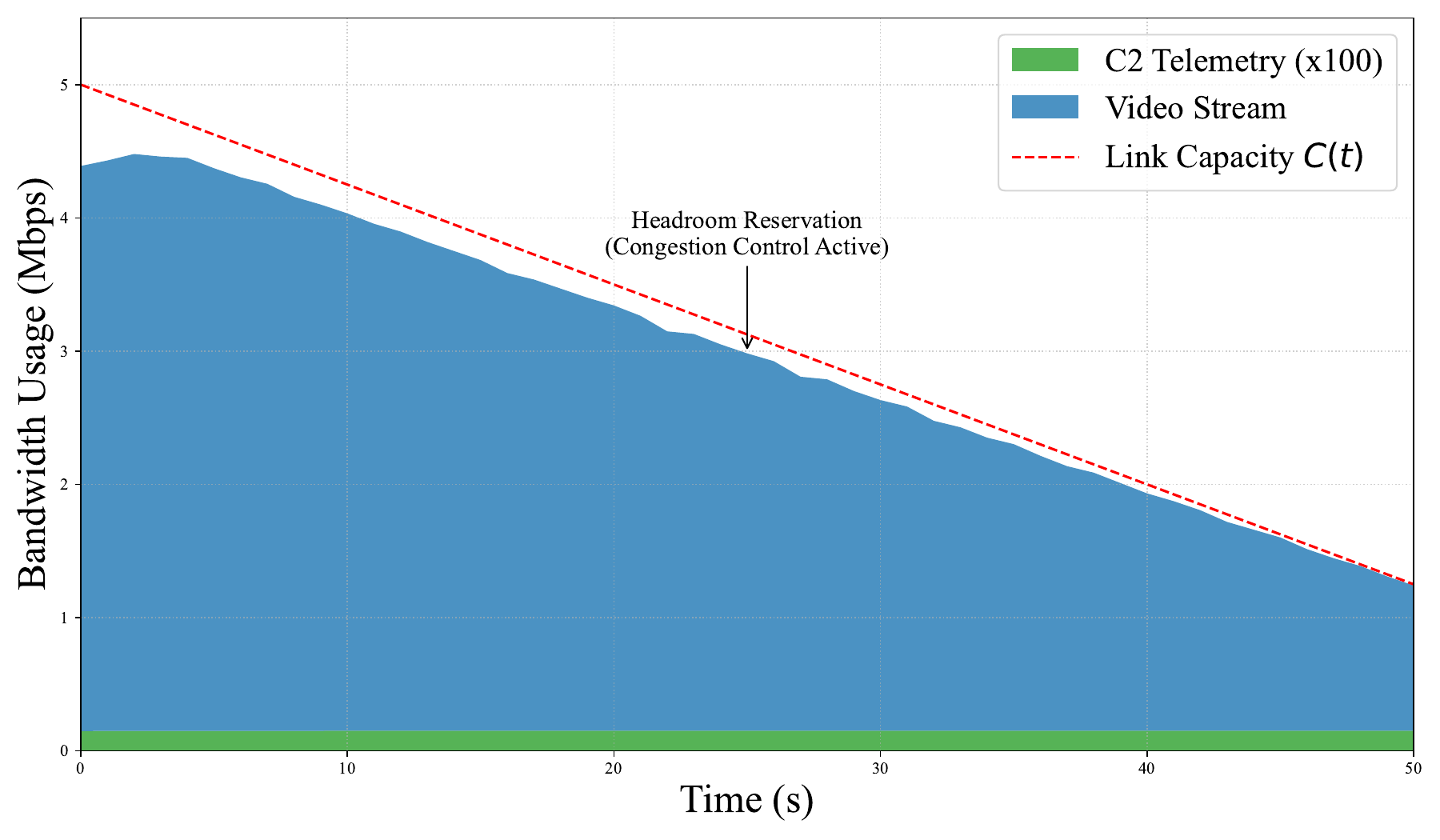}
    \caption{Efficacy of Cross-Layer Congestion Control under Decreasing Bandwidth. \textnormal{The stacked area chart depicts the real-time throughput of the prioritized C2 stream (bottom/dark) and the adaptive video stream (top/light) against a fluctuating physical link capacity $C(t)$ (dashed line). When the bandwidth drops from 5 Mbps to 1 Mbps, the video encoding rate $R_{enc}$ is dynamically throttled to strictly maintain a safety margin $R_{safe}$, as defined in Eq.~\ref{eq:rate_coupling}. This ensures that critical telemetry is never blocked by video data.}}
    \label{fig:headroom}
\end{figure}

Figure \ref{fig:headroom} visualizes the throughput dynamics of the scheduler. During the initial steady state (0-10s) where the link capacity is 5 Mbps, the video stream naturally consumes the majority of the available bandwidth, while the C2 traffic occupies a negligible but constant fraction. As the bandwidth $C(t)$ drops instantaneously to 1 Mbps at $t=10s$, the unified QUIC congestion controller detects the sharp increase in queuing delay via the estimator $\hat{d}_{q}(t)$. Unlike independent UDP streams which would continue to flood the link, AQUILA's transport layer atomically reduces the sending rate.

Crucially, the system prevents the video stream from consuming the full 1 Mbps capacity even after the drop. As observed in Figure \ref{fig:headroom}, a persistent gap remains between the aggregate throughput and the physical limit (represented by the dashed line). This gap corresponds to the \textit{Telemetry Headroom} ($R_{safe}$), reserved specifically to ensure that the reliable C2 stream never encounters a full buffer. This behavior empirically validates that the expected waiting time for C2 packets remains decoupled from the video traffic intensity, effectively eliminating the Head-of-Line blocking phenomena characteristic of traditional hybrid TCP/UDP architectures. Even under severe constraint, the C2 link retains a dedicated "fast lane," preventing the self-inflicted Denial-of-Service attacks described in Section 1.

\subsection{Handover Resilience and 0-RTT Reconnection}
\label{sec:0rtt_results}

To empirically validate the "Control Blackout Window" minimization analysis presented in Section 3.1 (Eq.~\ref{eq:minimize_window}), we conducted a series of reconnection stress tests. The objective was to measure the time elapsed between the initiation of a network handover and the restoration of a fully encrypted application data flow. We compared AQUILA's 0-RTT implementation against a standard TCP/TLS 1.3 stack under two distinct round-trip delay scenarios: a low-latency environment (20ms) typical of strong cellular signal, and a high-latency environment (120ms) representative of cell-edge conditions or satellite backhaul.

\begin{figure}[thpb]
    \centering
    \includegraphics[width=\linewidth]{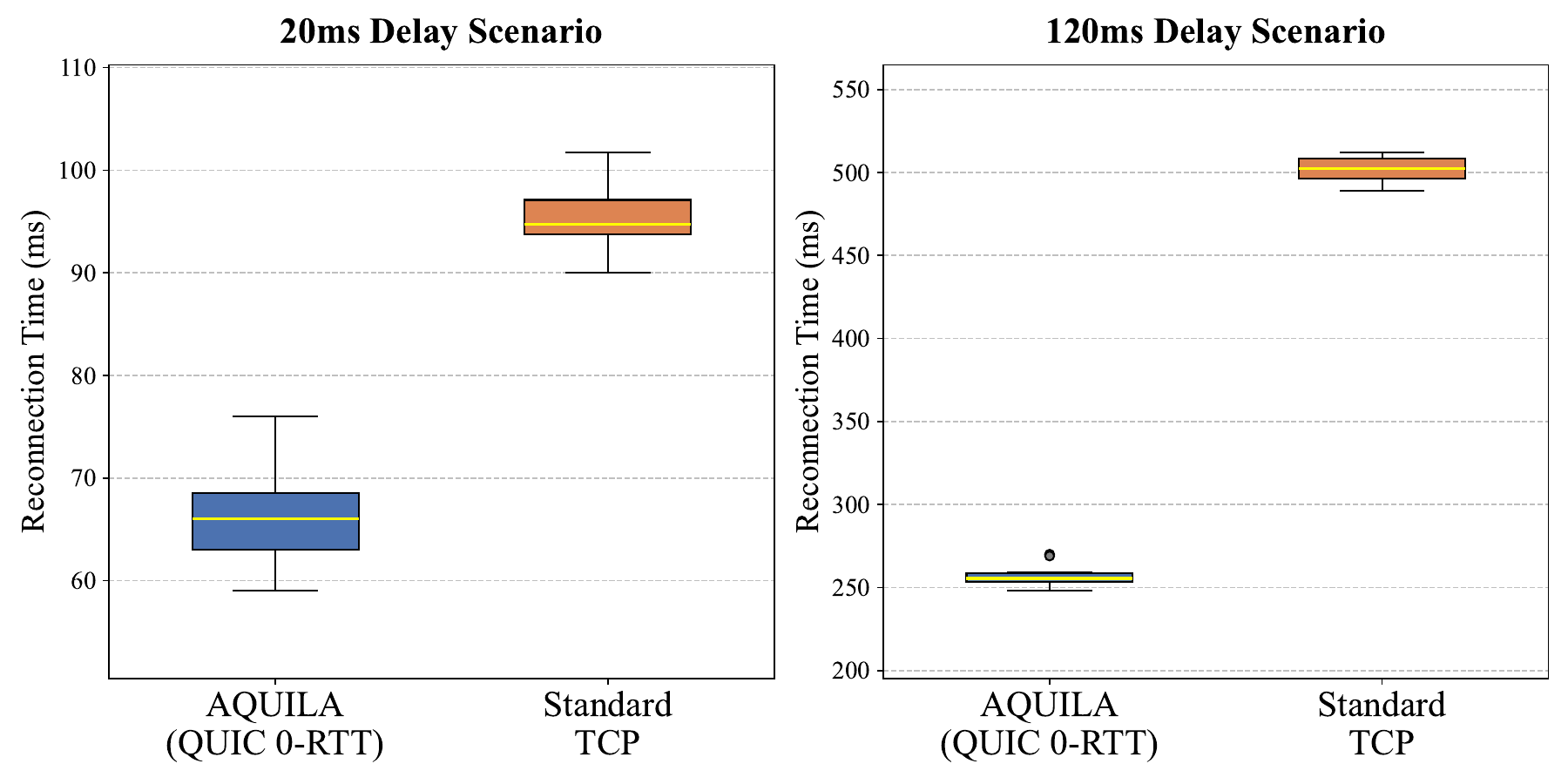}
    \caption{Impact of 0-RTT Session Resumption on Link Recovery Time. \textnormal{Comparison of reconnection latency between AQUILA (QUIC 0-RTT) and Standard TCP/TLS under simulated environment delays of 20ms and 120ms. AQUILA significantly reduces the control blackout window ($W_{cb}$) by eliminating the handshake overhead.}}
    \label{fig:0rtt}
\end{figure}

The results, visualized in Figure \ref{fig:0rtt}, demonstrate the substantial latency advantage of the QUIC 0-RTT mechanism.

\paragraph{High-Latency Scenarios (120ms):} The performance gap is most pronounced in high-latency environments. As shown in the box plot, the Standard TCP connection requires an average of 501ms ($\sigma \approx 7.2$) to re-establish the link. This delay stems from the mandatory multi-step handshake (TCP SYN/ACK followed by TLS key exchange), which necessitates multiple round trips before payload transmission can resume. In contrast, AQUILA achieves a mean reconnection time of 256ms ($\sigma \approx 6.3$). This $\approx 49\%$ reduction confirms that AQUILA effectively eliminates the protocol negotiation overhead ($T_{handshake} \approx 0$), leaving the reconnection time dominated primarily by the physical round-trip time ($2 \times RTT$ for confirmation) rather than protocol semantics.

\paragraph{Low-Latency Scenarios (20ms):} Even in the 20ms environment, AQUILA maintains a distinct advantage, with a mean recovery time of 65ms compared to 95ms for TCP. While the absolute difference is smaller, the relative reduction of $\approx 31\%$ remains critical for high-speed aerial maneuvers where control loop stability is sensitive to even minor jitter.

These findings validate that AQUILA's 0-RTT capability effectively decouples the control blackout duration from the protocol overhead. By ensuring that the reconnection time scales linearly with physical path latency rather than the number of handshake round-trips, AQUILA provides a safety-critical margin against the frequent handovers inherent in BVLOS cellular operations.

\subsection{Computational Overhead Analysis}
\label{subsec:overhead}

\begin{figure}[thpb]
    \centering
    \includegraphics[width=\linewidth]{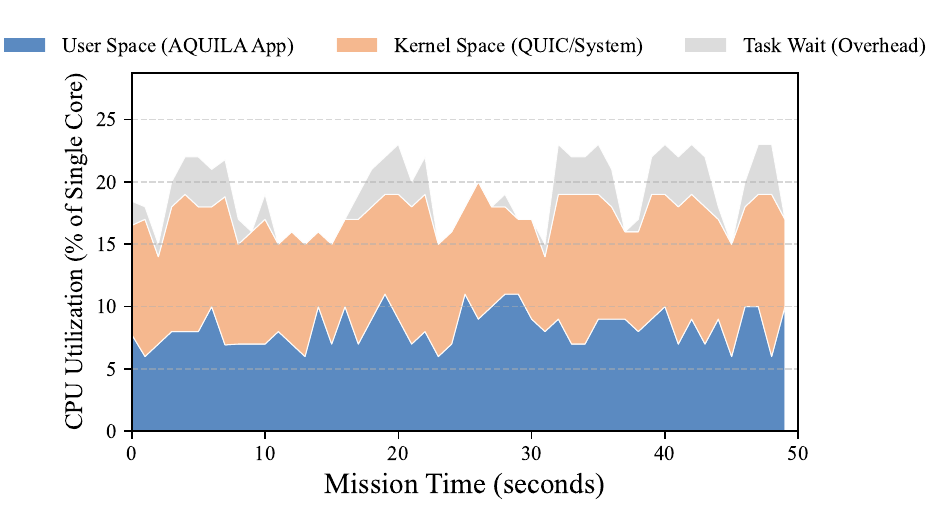}
    \caption{CPU Utilization of AQUILA Server on Onboard Computer. \textnormal{The stacked area chart depicts the resource consumption over a 50s transmission window. The \textit{User Space} component represents the AQUILA Rust application handling QUIC encryption and scheduling, while \textit{Kernel Space} accounts for UDP I/O. The low overhead demonstrates feasibility on embedded hardware.}}
    \label{fig:cpu_overhead}
\end{figure}

Given that the AQUILA architecture is deployed on a resource-constrained companion computer (Orange Pi 3B), demonstrating computational feasibility is as critical as verifying network performance. A primary concern with migrating the transport stack from the kernel (TCP) to user space (QUIC) is the potential increase in CPU cycles due to context switching and software-based encryption. To quantify this overhead, we profiled the \textit{Aquila Server} process during a continuous 1080p video transmission session. Figure~\ref{fig:cpu_overhead} presents the CPU utilization profile, decomposed into user space (\texttt{\%usr}), kernel space (\texttt{\%system}), and scheduler wait time (\texttt{\%wait}). The analysis yields critical observations regarding the efficiency of our Rust-based implementation. Throughout the 50-second measurement window, the total single-core CPU utilization consistently fluctuated between 14\% and 19\%. Considering the Orange Pi 3B is equipped with a quad-core Cortex-A55 processor, the AQUILA stack effectively consumes less than 5\% of the total available onboard computing capacity, leaving ample headroom for concurrent execution of flight control MAVLink forwarding and potentially computationally intensive tasks such as obstacle avoidance. Contrary to typical user-space applications, the profile shows that kernel-space operations (\texttt{\%system}, avg. $\approx$ 9--11\%) slightly exceed user-space processing (\texttt{\%usr}, avg. $\approx$ 6--8\%), indicating that the computational cost of AQUILA's core logic---including the Rust-based packet scheduling and encryption---is highly optimized. The dominant overhead stems from the syscalls required to push high-frequency UDP datagrams to the network interface, confirming that the choice of Rust and \texttt{quiche} effectively mitigates the performance penalty often associated with application-layer protocols. Furthermore, the CPU wait time (\texttt{\%wait}) remains negligible, peaking at only 4\% and averaging below 2\%, which suggests that the asynchronous runtime used for the priority scheduler is not inducing thread starvation. In summary, AQUILA demonstrates that modern, encrypted, user-space transport protocols can operate efficiently on low-power embedded hardware without compromising the SWaP (Size, Weight, and Power) constraints of small UAV platforms.

\section{RELATED WORK}

\subsection{Cellular-Enabled UAV Communication}
Research into Beyond Visual Line-of-Sight (BVLOS) operations has increasingly prioritized cellular networks to overcome the range limitations inherent in direct point-to-point radio links~\cite{8660516}. Extensive channel modeling studies have characterized the unique physical layer conditions faced by aerial platforms, identifying that Unmanned Aerial Vehicles (UAVs) at altitude frequently connect via the side lobes of base station antennas which are optimized for terrestrial coverage~\cite{lte_for_uav, 9473547}. Literature in this domain documents how this side-lobe connectivity contributes to signal volatility and inter-cell interference, leading to handover behaviors that differ significantly from ground-based mobility patterns. To address these connectivity challenges, the research community has proposed various architectural solutions, ranging from multi-link bonding techniques that aggregate bandwidth across providers~\cite{uav_multi_link} to heterogeneous network switching strategies that toggle between satellite and cellular bands~\cite{9275613}. While hardware-centric approaches focus on redundancy to ensure resilience, parallel software-defined research explores transport layer optimizations~\cite{cross_layer_uav} designed to maximize the stability of single cellular links within the strict size, weight, and power constraints of small aerial vehicles~\cite{skycore}.

\subsection{Transport Protocols for Real-Time Media}
The requirements of robotic teleoperation have driven significant investigation into mixed-criticality scheduling, where high-bandwidth video and low-latency control telemetry must coexist on a single link. Traditional networking architectures have typically addressed this by combining TCP for reliable command delivery with UDP for real-time video streaming~\cite{cross_layer_uav}, a pairing that has been the subject of extensive performance analysis regarding channel state awareness and synchronization. Beyond these hybrid approaches, the research community has evaluated established real-time transport standards. The Real-time Transport Protocol (RTP)~\cite{rtp} is widely utilized for media streaming, often augmented with Forward Error Correction (FEC) extensions~\cite{rtp_fec} to recover from packet loss without retransmission delays. Secure Reliable Transport (SRT)~\cite{srt, live_with_srt} has similarly been adopted for its ability to optimize performance over unpredictable networks. WebRTC~\cite{w3c-webrtc, rfc8825} has also attracted attention for its robust, built-in congestion control and security, employing bitrate adaptation schemes~\cite{bit_adapt} that couple transport feedback with the video encoder and dynamically align data rates with channel capacity to mitigate volatility. In the pursuit of lower latency, lightweight user-space protocols such as KCP~\cite{kcp_protocol} have been implemented to provide reliable delivery with aggressive retransmission strategies distinct from standard TCP dynamics. Consequently, recent studies have shifted focus toward modern transport protocols like QUIC~\cite{iyengar2021quic}, which move the transport stack into user space to support multi-streaming and eliminate Head-of-Line blocking between independent data flows. The standardization of unreliable datagram extensions~\cite{pauly2022unreliable} has further enabled researchers to explore unified transport architectures that handle both reliable control streams and delay-sensitive media within a single encrypted context.

\subsection{Congestion Control for Aerial Mobility}
The development of congestion control algorithms suitable for aerial environments remains a distinct area of study, with a primary focus on differentiating network congestion from wireless channel variability~\cite{cc_for_rtp}. While general-purpose algorithms like CUBIC~\cite{cubic} and BBR~\cite{bbr} utilize packet loss or model-based estimates for bulk data transfer, multimedia-specific research has yielded delay-based controllers such as Google Congestion Control (GCC)~\cite{gcc}, Network Assisted Dynamic Adaptation (NADA)~\cite{nada}, and Self-Clocked Rate Adaptation for Multimedia (SCREAM)~\cite{RFC8298}. These algorithms have been characterized by their use of delay gradients to minimize queue occupation and their respective trade-offs between convergence speed and bandwidth utilization. Building on these foundations, recent work has begun to address the vertical dimension of cellular networks, investigating altitude-aware modifications that can distinguish between bufferbloat and the structural latency spikes caused by handovers~\cite{handover}. These specialized controllers aim to maintain video quality while preserving necessary bandwidth specifically for safety-critical telemetry during transient link instability.

\section{DISCUSSION}

\paragraph{Architectural Trade-off: Determinism vs. Fairness}
The experimental results demonstrate that AQUILA effectively eliminates head-of-line blocking and maintains link integrity under dynamic conditions, but this performance stems from a fundamental architectural paradigm shift: the prioritization of safety determinism over network fairness. In traditional hybrid architectures, TCP and UDP flows compete for bandwidth as independent entities, often resulting in "context-blind" video streams that inadvertently flood the network buffers and starve the critical command and control (C2) link during congestion events~\cite{uav_comm_challenges}. AQUILA rejects this fairness model by establishing a strict hierarchy where video throughput is subservient to telemetry stability. This prioritization is mechanically enforced through the proposed Safety-Critical Rate Coupling, which dynamically throttles the video bitrate to maintain a reserved bandwidth margin, denoted as $R_{safe}$. While this strategy provides a mathematical guarantee that C2 latency remains bounded regardless of video traffic intensity , it introduces a deliberate trade-off: the system may prevent the video stream from utilizing the theoretical maximum link capacity to preserve this safety buffer. Consequently, peak video throughput is effectively sacrificed to guarantee absolute flight control authority, a necessary compromise for safety-critical BVLOS operations.

\paragraph{Limitation: Security Dependency of 0-RTT}
A significant limitation of the proposed architecture lies in the security implications of the 0-RTT connection resumption mechanism employed to handle cellular handovers. While AQUILA successfully minimizes the "control blackout window" by eliminating the cryptographic handshake overhead—allowing the reconnection time to approximate the physical round-trip time ($T_{handshake} \approx 0$)—this speed optimization removes forward secrecy for the initial data packet, rendering the transport layer theoretically susceptible to replay attacks~\cite{security_uav}. To mitigate this vulnerability, the system shifts the responsibility for replay protection from the encrypted transport stack to the application layer. The architecture relies on the MAVLink protocol’s embedded monotonic sequence numbers and timestamps to identify and silently discard duplicate commands, ensuring that replayed packets do not affect the flight controller. This approach creates a strong coupling between layers; the transport layer is not "secure by default" for generic payloads in 0-RTT mode and depends entirely on the upper-layer protocol to implement strict idempotency or sequence validation, thereby limiting the architecture's modularity for applications that lack these specific integrity checks. We plan to continue investigating this trade-off by characterizing the theoretical boundaries of 0-RTT security and exploring adaptive strategies that dynamically balance risk against the urgency of connection resumption.

\paragraph{Future Work}
Future iterations of AQUILA will address the rigidity of the current congestion control implementation and the vulnerability inherent in single-link operations. Currently, the modified SCReAM algorithm relies on static configuration parameters, such as the 20-second observation window and the tolerance factor ($\gamma=1.5$), which were heuristically tuned based on specific suburban LTE field tests. While effective in the tested environment, these static values may not generalize well to different cellular topologies or highly dynamic interference patterns. We plan to replace this heuristic tuning with an Adaptive Congestion Control system driven by Reinforcement Learning (RL)~\cite{shahid2022uavs, zhang2024advancing}. By training an agent to interpret real-time variations in Reference Signal Received Power (RSRP) and Signal-to-Noise Ratio (SNR) , the system could dynamically adjust delay targets and window sizes, allowing it to distinguish between transient channel noise and persistent congestion with greater accuracy than the current fixed-logic estimator.

Furthermore, to mitigate the risk of physical blind spots and signal degradation associated with a single cellular interface, we intend to extend the architecture to support Multipath QUIC (MP-QUIC). The current implementation relies on a single 4G/LTE modem, leaving the system vulnerable to coverage gaps or total signal loss in specific geographic areas. Adopting MP-QUIC would allow the UAV to aggregate bandwidth and seamlessly failover between heterogeneous links, such as bonding the existing LTE connection with satellite backhaul or secondary cellular carriers. This evolution would eliminate the single point of failure at the physical layer, ensuring that the unified transport context and priority scheduling mechanisms remain effective even if one underlying link degrades completely.
\section{Conclusion}

This work presented AQUILA, a cross-layer communication framework designed to secure the reliability of long-range unmanned aerial vehicle operations over cellular networks. By leveraging the multiplexing capabilities of the QUIC protocol, the proposed architecture establishes a unified transport layer that concurrently manages high-throughput video streams and latency-sensitive command telemetry within a single congestion control context. The integration of a structurally ensured priority scheduler, coupled with a modified congestion control algorithm featuring adaptive delay targeting and telemetry headroom reservation, effectively resolves the resource conflict between visual situational awareness and flight control stability. Furthermore, the implementation of zero-round-trip time session resumption significantly mitigates control blackouts during network handovers. Experimental validation confirms that this approach eliminates head-of-line blocking and ensures continuous link integrity under dynamic aerial channel conditions, offering a robust foundation for safe beyond visual line-of-sight missions.

\appendix
\section{Hardware and Software Specifications}

\begin{table}[H]
\begin{center}
\begin{tabular}{|l||l|}
\hline
\textbf{Component} & \textbf{Specification} \\
\hline
Airframe & Custom Fixed-Wing, 1100mm Wingspan \\
Flight Controller & Speedybee F405 Wing Mini \\
Onboard Computer & Orange Pi 3B (Rockchip RK3566) \\
4G/LTE Modem & Quectel EC20CEFAG \\
Telemetry Radio & RTL8822CU (5.8 GHz) \\
Camera & OpenIPC(GK7205v200 IMX307) \\
Motor & SunnySky X2212 980kv \\
ESC & Hobbywing Skywalker V2 40A \\
Servo & EMAX ES08MD \\
Airspeed Sensor & MS4525D \\
\hline
\end{tabular}
\end{center}
\caption{UAV Hardware Specifications}
\label{table_hardware}
\end{table}

\begin{table}[H]
\begin{center}
\begin{tabular}{|l||l|}
\hline
\textbf{Software} & \textbf{Specification} \\
\hline
AQUILA & github.com/Vinylether/aquila \\
Quiche & v0.22 (cargo 1.91.1) \\
ffmpeg & 8.0.1 w/ libvmaf \\
GStreamer & v1.22.0 \\
Mahimahi & git commit f1346c3 (2025-04-15) \\
ArduPilot & v4.6.2 \\
MAVLink & v1.0.12 \\
Onboard Computer System & Debian, Kernel 6.6.0-rc5 \\
Azure Instance System & Rocky Linux 9.5, Kernel 5.14.0 \\
\hline
\end{tabular}
\end{center}
\caption{UAV and Experimental Software Specifications}
\label{table_software}
\end{table}

\section{Experimental Data}

\begin{figure}[H]
    \centering
    \includegraphics[width=\linewidth]{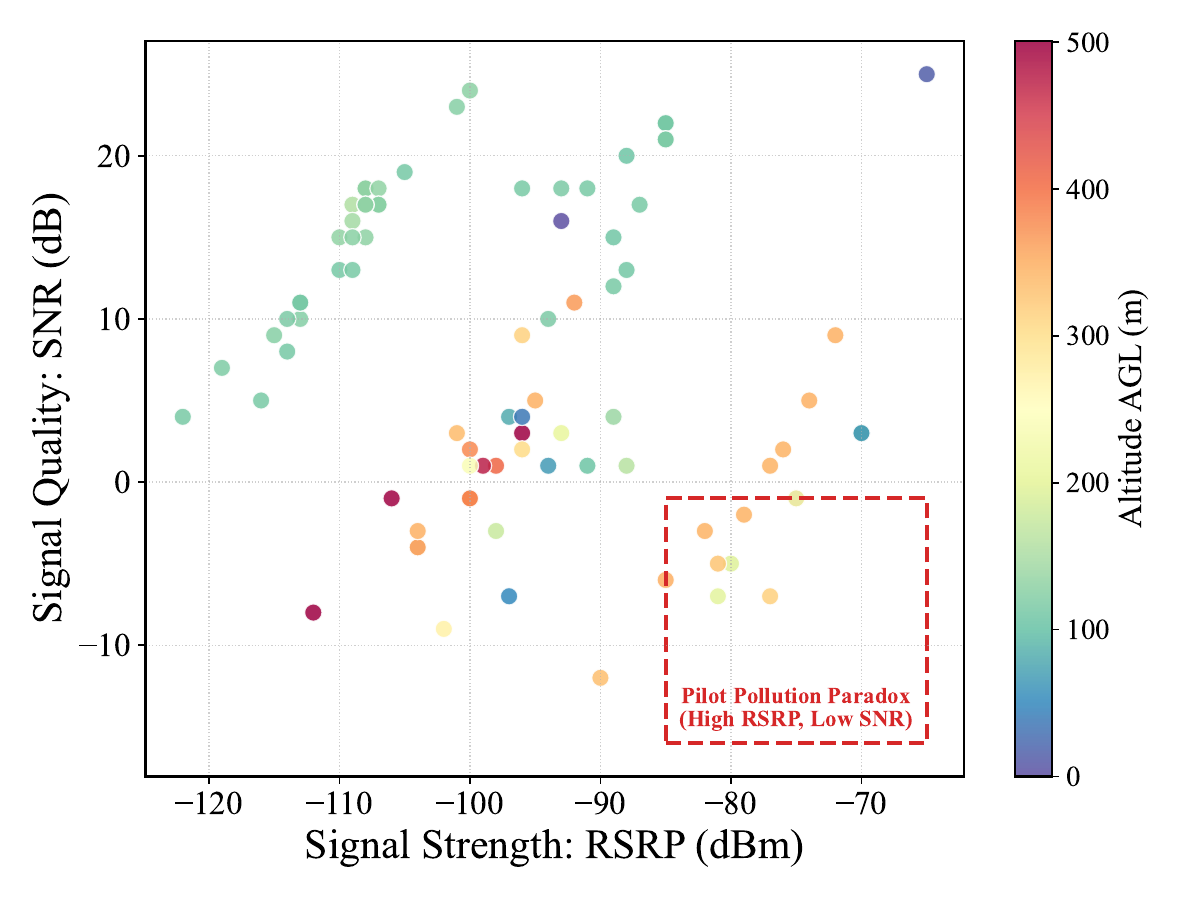}
    \caption{Pilot Pollution at High Altitude: Strong Signal, Poor Quality}
    \label{fig:pilot_pollution}
\end{figure}

\begin{figure}[htbp]
    \centering

    \begin{subfigure}[b]{\linewidth}
        \centering
        \includegraphics[width=\linewidth]{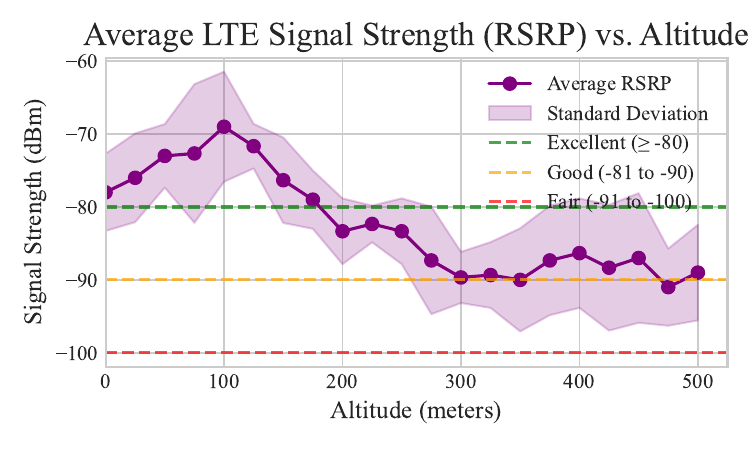}
    \end{subfigure}

    \vspace{0.3cm}
        
    \begin{subfigure}[b]{\linewidth}
        \centering
        \includegraphics[width=\linewidth]{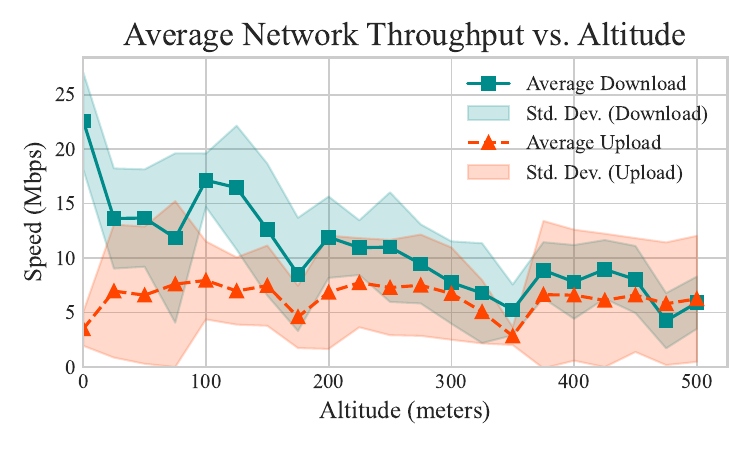}
    \end{subfigure}
    
    \vspace{0.3cm}

    \begin{subfigure}[b]{\linewidth}
        \centering
        \includegraphics[width=\linewidth]{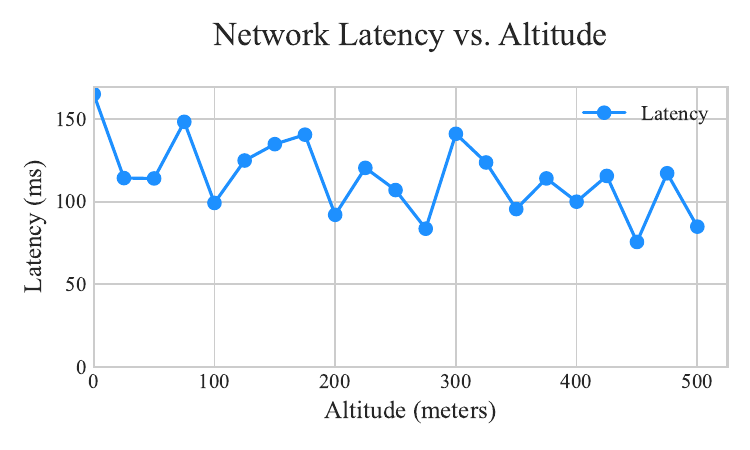}
    \end{subfigure}

    \vspace{0.3cm} 

    \begin{subfigure}[b]{\linewidth}
        \centering
        \includegraphics[width=\linewidth]{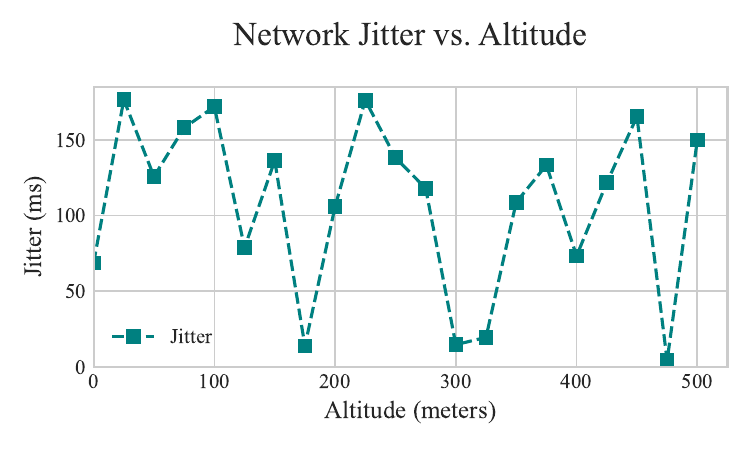}
    \end{subfigure}

    \caption{Field Test Data.}
    \label{fig:fieldtest}
\end{figure}

\bibliographystyle{unsrt}
\bibliography{references}

\end{document}